\documentclass[prd,a4paper,twocolumn,preprintnumbers,nofootinbib,superscriptaddress]{revtex4}

\pdfoutput=1
\usepackage[english]{babel}
\usepackage{amsmath,amssymb,amsfonts, bm,bbm,slashed, subdepth}
\usepackage{graphicx}
\usepackage[colorlinks=true]{hyperref}
\usepackage{cleveref}
\usepackage{enumerate}
\usepackage{setspace}
\usepackage{booktabs, tabularx}
\usepackage{units}
\usepackage{slashed}
\usepackage[dvipsnames]{xcolor}
\usepackage{soul}
\usepackage{multirow}

\usepackage{geometry}
\newcommand{\eV}{\mbox{\rm eV}}

\newcommand{\TeV}{\mbox{\rm TeV}}

\newcommand{\diff}{\mathrm{d}}

\makeindex

\begin{document}
\preprint{ULB-TH/21-05}
	
	\title{The domain of thermal dark matter candidates
	}

	\author{Rupert Coy}
	\email{rupert.coy@ulb.ac.be}
	\affiliation{Service de Physique Th\'eorique, Universit\'e Libre de Bruxelles, Boulevard du Triomphe, CP225, 1050 Brussels, Belgium}
	
	\author{Thomas~Hambye}
	\email{thambye@ulb.ac.be}
	\affiliation{Service de Physique Th\'eorique, Universit\'e Libre de Bruxelles, Boulevard du Triomphe, CP225, 1050 Brussels, Belgium}

	\author{Michel H.G. Tytgat}
	\email{michel.tytgat@ulb.ac.be}
	\affiliation{Service de Physique Th\'eorique, Universit\'e Libre de Bruxelles, Boulevard du Triomphe, CP225, 1050 Brussels, Belgium}

	\author{Laurent Vanderheyden}
	\email{lavdheyd@ulb.ac.be}
	\affiliation{Service de Physique Th\'eorique, Universit\'e Libre de Bruxelles, Boulevard du Triomphe, CP225, 1050 Brussels, Belgium}
	
	\begin{abstract}
		We consider, in general terms, the possible parameter space  of {thermal} dark matter candidates.  {We assume} that the dark matter particle is fundamental and was in thermal equilibrium in a hidden sector with a temperature $T'$, which may differ from that of the Standard Model {temperature}, $T$. 
		{The candidates} lie in a region in the $T'/T$ vs. $m_{\rm dm}$ plane, which is bounded by both {model-independent} theoretical considerations and observational constraints. 
		The former consists of limits from dark matter candidates that decoupled when relativistic ({the} relativistic floor) and {from} those that decoupled when non-relativistic with the largest annihilation cross section allowed by unitarity ({the} unitarity wall), while the latter concerns big bang nucleosynthesis ($N_{\rm eff}$ ceiling) and free streaming.
		{We present three simplified dark matter scenarios, demonstrating concretely how each fits into the domain.}
	\end{abstract}
	
	\maketitle

	\section{Introduction}
	Of all the scenarios for dark matter (DM), one of the {most} appealing possibilities is that it is made of elementary particles that were in thermal equilibrium in the early universe. If this is the case, then just like the photons of the cosmic microwave background, their abundance today reflects the one they had when they decoupled, or `froze out', from their thermal environment \cite{Cowsik:1972gh,Lee:1977ua,Kolb:1990vq}. We refer to DM candidates that were once in thermodynamic equilibrium as thermal candidates.
	One class of thermal candidates consists of particles that decoupled when they were non-relativistic. Because of the exponential dependence of their Maxwell-Boltzmann distribution, the moment of freeze-out is critical: the later the decoupling, the smaller their relic abundance and {\em vice versa}. Alternatively, the DM particles may have decoupled when they were relativistic. In that case, their relic abundance depends on the moment of decoupling only through entropy transfer effects (like {for the Standard} Model (SM) relic neutrinos). 
	The primary question we would like to address in this work is the following: what is the domain of all possible thermal candidates?

	The answer is well known if the DM particles were in thermal equilibrium with SM particles. As we will review in section \ref{sec:1D}, thermal candidates {which give the correct relic abundance} lie {somewhere between a few $\eV$ (a value which is ruled out by structure formation constraints, as is well known) and about} 100 $\TeV$. 
	The lower bound corresponds to DM that decoupled when relativistic \cite{Cowsik:1972gh}, and the upper bound to DM that decoupled when non-relativistic with the largest possible annihilation cross section compatible with unitarity \cite{Griest:1989wd}. But what if the DM is secluded and lives in a hidden sector (HS) with a temperature, $T'$, that differs from that of the thermal bath of SM particles (the visible sector, or VS), which has temperature $T$? 
	The possibility that DM resides in a HS (see e.g. \cite{Pospelov:2007mp,Feng:2008mu,Chu:2011be}), possibly with its own temperature, perhaps due to some reheating story after inflation (see e.g. \cite{Hodges:1993yb,Berezhiani:1995am}), has by now been studied in many scenarios. However 
	the {full} domain of possibilities has not been systematically discussed in the literature. In this work, we aim to determine it in general.
	This domain of all DM thermal candidates is two dimensional, as it depends both on the DM mass and on the temperature ratio, $\xi = T'/T$. 
	In particular, we will derive the equivalent of the lower and upper mass bounds on thermal DM candidates as a function of $T'/T$. 
	We will also determine which part of this domain is excluded by various observational constraints.
	
	The plan of this paper is as follows. To lay the groundwork, we will begin in section \ref{sec:1D} by briefly reviewing the standard result for particles that were in thermal equilibrium with the SM. In section \ref{sec:2D}, we will extend this to DM particles which reside in a HS. In section \ref{sec:theory}, we will discuss the theoretical constraints, in particular from unitarity. Next, in section \ref{sec:exp}, we will discuss the bounds from limits on the number of relativistic degrees of freedom at the time of big bang nucleosynthesis (BBN) on one hand, and from free-streaming (FS) on the other. 
	The main result of our analysis is summarised in Fig.~\ref{fig:polygon}, which depicts the domain (white region) of all possible thermal candidates which survive these constraints in the plane $T'/T$ {\em vs.} $m_{\rm dm}$. The temperature ratio here is defined at the moment of decoupling of the DM particles. Three simple but illustrative explicit DM models are considered in section \ref{sec:models}. We draw our conclusions in section \ref{sec:conclusions}. This work is built upon and extends results from \cite{Hambye_2020,Hambye:2020lvy}. 
	
	\section{Domain for $T'=T$}
	\label{sec:1D}
	We start by reviewing a classic result regarding possible thermal candidates that were in thermal equilibrium with the SM thermal bath. If this were the case, the lightest {possible} thermal candidate would have a mass at the eV scale. 
	This lower bound is obtained assuming the largest possible number of DM particles when it decouples, that is to say when DM decouples relativistically,
	in which case
	\begin{eqnarray}
		\Omega_{\rm dm} h^2 &=& \frac{ m_{\rm dm} n_{\rm dm,0}}{\rho_{c,0}/h^2} \equiv {m_{\rm dm} Y_{\rm dm,dec} s_0 \over \rho_{c,0}/h^2} \nonumber\\
		&\approx&  0.12 \,\left({g' m_{\rm dm}\over 6 \,{\rm eV}}\right) \left({g_{\ast s,0}\over g_{\ast s,{\rm dec}}}\right)\,.
		\label{eq:relabundance} 
	\end{eqnarray}
	The parameter $g' = c \cdot g_{\rm dm}$ counts the effective degeneracy of the DM candidate, where $c$ is 1 (3/4) for a boson (fermion) and $g_{\rm dm}$ is the internal DM degrees of freedom, while $g_{\ast s,0}$ counts the total number of degrees of freedom contributing to the entropy today, $g_{\ast s,{\rm dec}}$ is the same but at DM decoupling and $\rho_{c,0}$ is the critical energy density today. From this, the DM mass is required to be at least
	\begin{equation}
		\label{eq:CM}
		m_{\rm dm} \gtrsim m_{\rm dm,CM} \approx {6\over g'} \left({g_{\ast s,{\rm dec}}\over g_{\ast s,0}}\right)\,\mbox{\rm eV} \,.
	\end{equation}
	In the context of SM neutrinos, Eq.~(\ref{eq:CM}) is the Cowsik-McClelland (CM) bound \cite{Cowsik:1972gh}.\footnote{For SM neutrinos, $g' \rightarrow g_{\rm \nu} = 2 \times 3/4$ per species and, due to entropy transfer from the electrons and positrons to the photons {after SM neutrino decoupling, $g_{\ast s,0}/g_{\ast s,\rm dec} =4/11$. For $\Omega h^2 \lesssim 1$, this gives $\sum m_{\nu} \lesssim 92$ eV. 
			Here, in} general, we have in mind that $g'$ is of the order of a few, corresponding to ${\cal O}(1)$ species of DM. See \cite{Davoudiasl:2020uig} for a recent scenario in which $g'$ is considered to be a huge number.} It is usually considered as an upper limit on the sum of neutrino masses.  In the present context, it is a lower limit since it implies that the DM abundance is too small when $m_{\rm dm} < m_{\rm dm,CM}$.
	
	A thermal candidate with a larger mass than (\ref{eq:CM}) must decouple when non-relativistic so that its abundance is Boltzmann suppressed. This is the WIMP (weakly interacting massive particle) scenario, in which the
	abundance is inversely proportional to the annihilation cross section (thermally averaged at the time of freeze-out), which is required to be about $\langle \sigma v\rangle \approx 3\cdot 10^{-26}$ cm$^3$/s to match observations of the DM abundance \cite{Aghanim:2018eyx}. Although this number depends on the DM mass only logarithmically, the cross-section itself depends on it, albeit in a model-dependent way. In the non-relativistic limit, relevant for WIMP freeze-out, and when the DM mass is the largest relevant mass scale,  {the annihilation cross section scales like} $\langle \sigma v\rangle \propto 1/m_{\rm dm}^2$. For fixed $m_{\rm dm}$, unitarity sets the upper limit on the DM mass by fixing the maximal possible annihilation cross section of a pair of particles  with angular momentum $J$,
	\begin{equation}
		\sigma v \leq {4 \pi (2 J +1)\over m_{\rm dm}^2 v} \,,
		\label{eq:unitarity}
	\end{equation}
	{where $v$ is their relative velocity.}
	This translates into an upper limit on the mass of thermal candidates, known as the Griest-Kamionkowski (GK) bound \cite{Griest:1989wd}, 
	\begin{equation}
		\label{eq:GK1D}
		{ m_{\rm dm } \lesssim \; } m_{\rm dm,GK} \approx {3}00\; \mbox{\rm TeV} \, .
	\end{equation}
	Together, the lower   \eqref{eq:CM} and upper \eqref{eq:GK1D} limits
	define a one-dimensional domain,
	\begin{equation}
		\label{eq:1D}
		m_{\rm dm} \in \left [m_{\rm dm,CM} ,  m_{\rm dm,GK}\right] \,.
	\end{equation}
	{Below}  this range, the DM is {always} under-abundant, {while above}  it is over-abundant. 
	The abundance of the DM increases with the DM mass at both the lower and upper limits in (\ref{eq:1D}) (with $\Omega_{\rm dm,CM} \propto m_{\rm dm}$ and $\Omega_{\rm dm,GK} \propto m_{\rm dm}^2$ {respectively}). Thus, assuming continuity of the abundance as a function of the DM mass, $\Omega_{\rm dm}(m_{\rm dm})$ implies that for any given DM model, one obtains the observed relic abundance for
	at least one value of $m_{\rm dm}$ within the range of Eq.~(\ref{eq:1D}).
	Where precisely the thermal candidates lie is model-dependent, so the scope of this continuity argument is limited (the DM mass range \eqref{eq:1D} spans about 13 orders of magnitude), but is robust.
	Another generic result is that for a given model 
	there is in general an odd number of DM candidates, except for specific and thus fine-tuned choices of parameters (this will be seen more explicitly when we consider specific models in section \ref{sec:models}). 
	
\begin{figure}
		\centering
		\includegraphics[height=7.5cm]{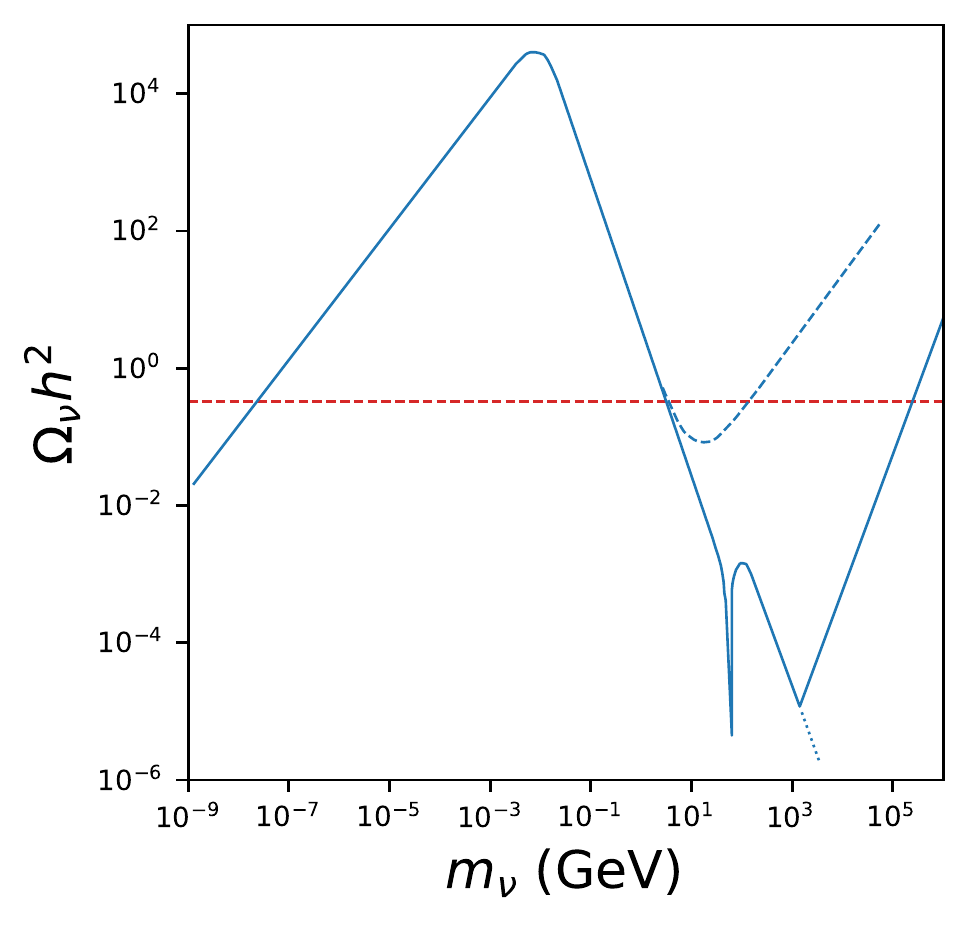}
		\caption{Relic density of Dirac neutrinos (solid). Adapted from \cite{Kainulainen:2002pu}, with courtesy of K. Kainulainen. 
			The solid (dashed) line corresponds to symmetric (asymmetric) DM. The dotted line is unphysical, as neutrinos become strongly interacting above $\sim$ few TeV. The red line corresponds to $\Omega_\nu h^2 = 0.12$. }\label{fig:neutrinos}
	\end{figure}

	These well known facts are  illustrated by the case of a massive Dirac neutrino \cite{Cowsik:1972gh,Lee:1977ua,Enqvist:1988we}. Consider Fig. \ref{fig:neutrinos}, reproduced from \cite{Kainulainen:2002pu}. It shows $\Omega_{\rm dm}h^2$ as a function of $m_{\rm dm}$, {\em i.e.} the neutrino mass, the only free parameter in this setup. As the figure shows, there are three possible thermal candidates: one at low mass, corresponding to the CM mass bound; a heavy one, corresponding to the GK mass bound\footnote{The behaviour of the relic abundance as a function of the neutrino mass  for masses above the $Z$ resonance is peculiar. The fact that it decreases is due to emission of longitudinal gauge bosons {with a cross-section that scales as $\langle \sigma v\rangle \propto m_\nu^2$, which therefore breaks} the unitarity limit for {sufficiently large} neutrino masses. {This} part of the curve that leads to the GK candidate is clearly schematic, see \cite{Kainulainen:2002pu}.}; and an intermediate one, with a mass in the GeV range. The latter is the Lee-Weinberg (LW) bound \cite{Lee:1977ua}. There are thus three possible candidates in this simple model (all of which are excluded by well known constraints). 
	
	We want to argue that all models with thermal DM candidates share similar features. 
	What are possible caveats? The most important hypothesis is that DM was in thermal equilibrium in the early universe. We will come back to this condition and its implications in section \ref{sec:therm}. 
	Here, we assume DM in thermal equilibrium and explore the other possible shortcomings, beginning with the GK mass bound. 
	The GK mass bound (\ref{eq:GK1D}) is quite general but still rests on some assumptions. First, it assumes that the DM particle annihilates with itself or with its antiparticle. Instead, it is possible that the abundance is set by co-annihilation of the DM particle with some almost degenerate companions \cite{Griest:1990kh,Gondolo:1990dk}. Provided there is sufficiently fast chemical equilibrium between these companion particles and the DM candidate, the DM abundance is determined by the most efficient annihilation channel(s). The GK mass bound thus applies to the companion particles, whose mass cannot be too much larger than the DM particle itself. 
	Hence $m_{\rm dm} \lesssim m_{\rm dm,GK}$ in that case too. 
	
	Another possibility is that DM is complex and that, like for the baryons in our universe, the DM abundance is set by an asymmetry instead of thermal freeze-out, {see e.g.~\cite{Zurek:2013wia,Petraki:2013wwa}. This} scenario is also subject to a GK bound. Indeed, a key feature of the asymmetric DM scenario is that the thermal, or symmetric, component of the DM abundance must be efficiently depleted, otherwise the asymmetry is hardly relevant. Efficient annihilation of the DM particle-antiparticle pairs (or their companions, as in the co-annihilation scenario) requires a larger cross section than the canonical value $3 \cdot 10^{-26}$ cm$^2$/s. This is suggested by the dashed line depicted in Fig.~\ref{fig:neutrinos}. In the absence of an asymmetry, the thermal abundance would be far below observations. Translating this to the maximum possible cross section set by unitarity considerations, the mass of the asymmetric DM must be below the standard GK bound (\ref{eq:GK1D}), 
	$m_{\rm dm,asym} < m_{\rm dm,GK}$.
	
	Finally, the GK mass bound only applies if DM particles are fundamental. Indeed, composite DM candidates \cite{Griest:1989wd} with radius much larger than their Compton wavelength, $r_{\rm dm}\gg 1/m_{\rm dm}$, can have a geometric cross section, $\sigma \sim \pi r_{\rm dm}^2$ (see, for instance, \cite{Harigaya:2016nlg} for a recent concrete example). With this possible exception kept in mind, we assume that generic thermal particle DM candidates all fall in the domain \eqref{eq:1D}.
	
	\section{Domain  for $T'\neq T$}
	\label{sec:2D}

\begin{figure}[b]
		\centering
		\includegraphics[height=7.5cm]{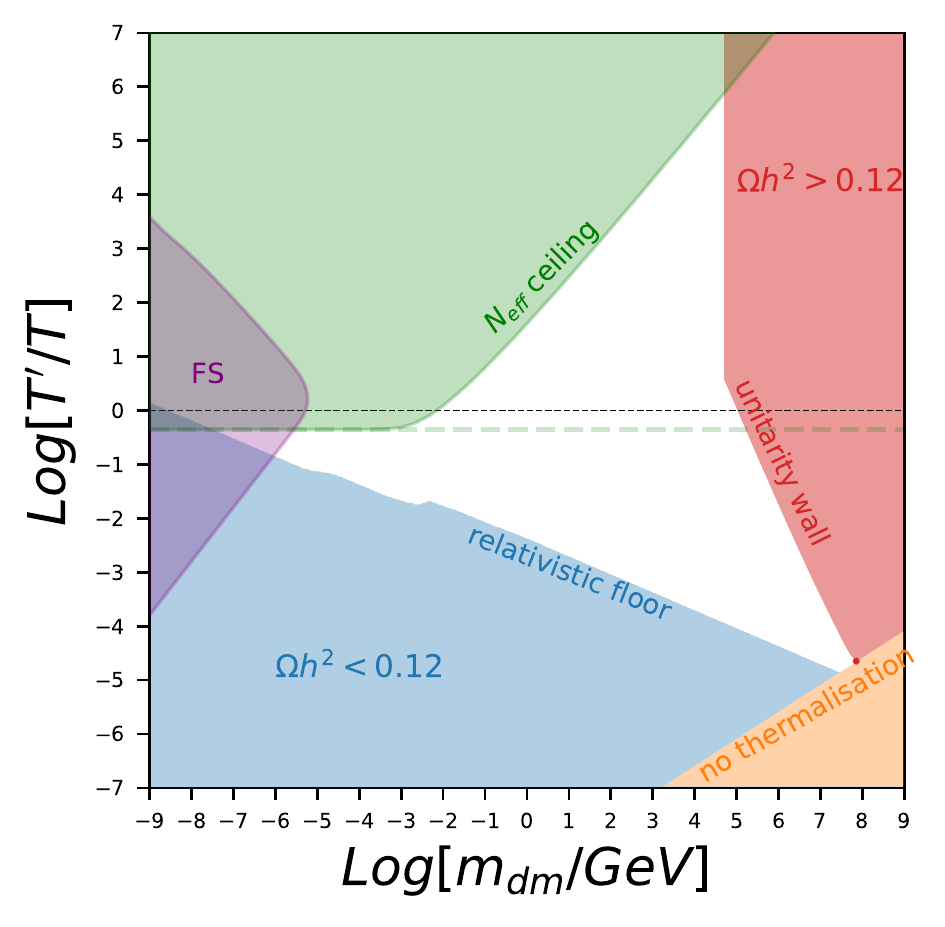}
		\caption{Domain of thermal DM candidates: temperature ratio $\xi$ at the time of DM decoupling vs. the DM mass. The possible thermal DM candidates lie in the white region. The coloured regions are excluded, see the text for how they are set. 
		}
		\label{fig:polygon}
	\end{figure}

	We now extend the established results of section \ref{sec:1D} to scenarios in which the DM was in thermal equilibrium in a hidden sector (HS). This HS is assumed to be  feebly interacting with the VS, so that the different sectors may have distinct temperatures, $T'\neq T$, at the moment of DM decoupling. Then, for a given $T'/T \neq 1$, we expect a generalised form of the interval (\ref{eq:1D}),
	with 
	\begin{equation}
		\label{eq:1Dxi}
		m_{\rm dm}(T'/T) \in \left [m_{\rm dm,CM}(T'/T) ,  m_{\rm dm,GK}(T'/T)\right]\,.
	\end{equation}
	Understanding the dependence of this range on $T'/T$ is one of our main goals. The result is summarised in Fig.~\ref{fig:polygon} (white region) in the plane $T'/T$ {\em vs.} $m_{\rm dm}$. The boundaries of that regions are explained in sections \ref{sec:theory} and \ref{sec:exp}.  Illustrations in terms of explicit models are given in section \ref{sec:models}.
	From now on, we use $\xi \equiv T'/T$ for the ratio of temperatures of the hidden and visible sectors. 
	
	\subsection{Theoretical bounds}
	\label{sec:theory}
	In this section, we consider theoretical constraints on the domain of all possible thermal DM candidates. These will be set by 1) the   DM abundance from relativistic decoupling (relativistic floor), 2) the condition of thermalisation of the DM within the HS (thermalisation condition), and 3) the DM abundance set from decoupling when non-relativistic with largest possible cross section (unitarity wall). We study the corresponding limits in that order. 
	
	\subsubsection{Relativistic floor}
	We first discuss the generalisation of the Cowsik-McClelland bound for $\xi \neq 1$.  The basic assumption is that the DM particle decoupled when relativistic at a temperature $T^\prime_{\rm dec}$. For given values of $m_{\rm dm}$ and $\xi$, relativistic decoupling generates the largest possible relic DM abundance. Comparing with the observed DM density gives a relation between $\xi$ (defined at decoupling) and the DM mass, $\xi(m_{\rm dm})$. Following \cite{Hambye_2020,Hambye:2020lvy}, we call this relation the relativistic floor as for fixed $m_{\rm dm}$, $\xi < \xi(m_{\rm dm})$ gives too little DM. 
	As in the previous section, the degeneracy of the DM species will be denoted by $g'$, including a factor $3/4$ for fermionic DM. The effective number of relativistic degrees of freedom in entropy in the HS is denoted with a prime, $g'_{\ast s}$. 
	
	One can distinguish two main scenarios of relativistic decoupling. The first scenario is analogous to the neutrino relativistic decoupling scenario in the SM.
	In the same way that the interaction rate for SM neutrinos changed from $\Gamma \propto \alpha_W^2 T$ to ${\Gamma} \propto G_F^2 T^5$ for $T \lesssim M_W$ \cite{Kolb:1990vq}, we can imagine that the DM interaction rate was $\sim T^{\prime 5}/\Lambda^4$ below some scale $T^\prime \sim \Lambda$, with $\Lambda {> T'_{\rm dec} } \gg m_{\rm dm}$. 
	Typically, $\Lambda$ is the mass of some heavy mediator. 
	The second scenario instead invokes a heavy ``cut-off" scale which is not the mass of a heavy mediator but the mass of the particles the DM scatters into. If these particles are heavier than the DM, the annihilation rate became Boltzmann suppressed when $T'$ went below their mass.
	Schematically, this is analogous to non-relativistic DM decoupling in the sense that decoupling was due to a Boltzmann suppression. 
	However, this happens at a higher temperature than $m_{\rm dm}$, when DM is still relativistic. 
	A difference between these two scenarios is that in the second one, DM is reheated just before its decoupling, as the heavier particle becomes non-relativistic and annihilates into it (similarly to photons which are reheated during the $e^+e^-$ annihilation catastrophe). Thus, in this case, the initial value of $T'/T$, which we call $\xi_{in}$ (before the heavier particle became non-relativistic), is different from the value of $T'/T$ at DM decoupling, $\xi_{dec}$. They differ by
	\begin{equation}
		\xi_{\rm dec}= \left({g'_{\ast s,{\rm in}} \over g'_{\ast s,\rm dec}} \right)^{1/3} \xi_{\rm in}
	\end{equation}
	where $g'_{\ast s,{\rm in}} > g'_{\ast s,dec}$ counts the HS degrees of freedom at the corresponding times.\footnote{As an example, consider a HS consisting of massive dark photons, electrons and positrons. The dark photon is the lightest particle of the HS, $m_{\gamma'} < m_{e'}$, and {so is a DM} candidate. In that case, $g'_{\ast s,\rm dec}$ counts the dark photon's degrees of freedom while $g'_{\ast s,{\rm in}}$ also includes the dark electrons and positrons. Also,  the initial dark photon  temperature corresponds to $T'_{\rm in} \approx m_{e'}$ so that the abundance of the dark photons after {decoupling is $Y_{\rm \gamma'} = (g'_{\ast s,{\rm in}}/g'_{\ast s,{\rm dec}}) Y_{\rm dm}(T'=m_{e'})$. More} generally, the dark photons could have inherited the entropy from all the charged particles that were once in equilibrium in the HS.} In the first scenario instead there is no such reheating of the HS and $\xi_{in}=\xi_{\rm dec}$ as $(g'_{\ast s,{\rm in}}/g'_{\ast s,{\rm dec}})=1$. 
	
	Taking into account this possible effect, we find the general formula
	\begin{eqnarray}
		\label{eq:relxi}
		\Omega_{\rm dm} h^2 &=& {m_{\rm dm} n_{\rm dm,0}\over \rho_{c,0}/h^2} \equiv {m_{\rm dm} Y_{\rm dm,dec} s_0 \over \rho_{c,0}/h^2}\\
		&=&  0.12 \, \left({g' m_{\rm dm} \over 6 \,{\rm eV}}\right)\left({g_{\ast s,0} \over g_{\ast s,{\rm dec}}} \right) \left({g'_{\ast s,{\rm in}} \over g'_{\ast s,{\rm dec}}} \right)\xi_{\rm in}^3\,.\nonumber
	\end{eqnarray}
	where $g_{\ast s,0}$ and $g_{\ast s,{\rm dec}}$ count the total of degrees of {freedom (i.e.~from both sectors)} at the corresponding times. 
	{Inversion of Eq. \eqref{eq:relxi} gives the $\xi_{\rm in}$ required to produce the correct DM abundance,
		\begin{equation}
			\xi_{\rm in} = \left(\frac{50 \,{\rm eV}} {g' m_{\rm dm}} \right)^{1/3} \left(\frac{g_{\ast s,{\rm dec}}}{g_{\ast s,0}} \right)^{1/3} \left(\frac{g'_{\ast s,{\rm dec}}}{g'_{\ast s,{\rm in}}} \right)^{1/3} \, .
			\label{eq:floor}
		\end{equation}
		This relation defines the relativistic floor, {giving the 
			blue} exclusion region in Fig. \ref{fig:polygon}. 
		The small kinks along the floor are due to the {evolution of the VS contribution to} $g_{\ast s}$. 
	}

	\subsubsection{Thermalisation condition}
	\label{sec:therm}
	By definition, a thermal DM candidate was in thermodynamic equilibrium. We must make sure that this condition was satisfied.
	In the following, we refer to thermal equilibrium as both kinetic and chemical equilibrium, so that the DM initial abundance is entirely determined by the HS temperature, $T'$.\footnote{Assuming DM chemical equilibrium is in general a stronger assumption, as processes that keep DM in kinetic equilibrium are generically more efficient, in particular when the DM is non-relativistic. Consequently, in the following, we use chemical equilibrium rates to impose the condition of thermal equilibrium.} 
	
\begin{figure}
		\centering
		\includegraphics[height=7.5cm]{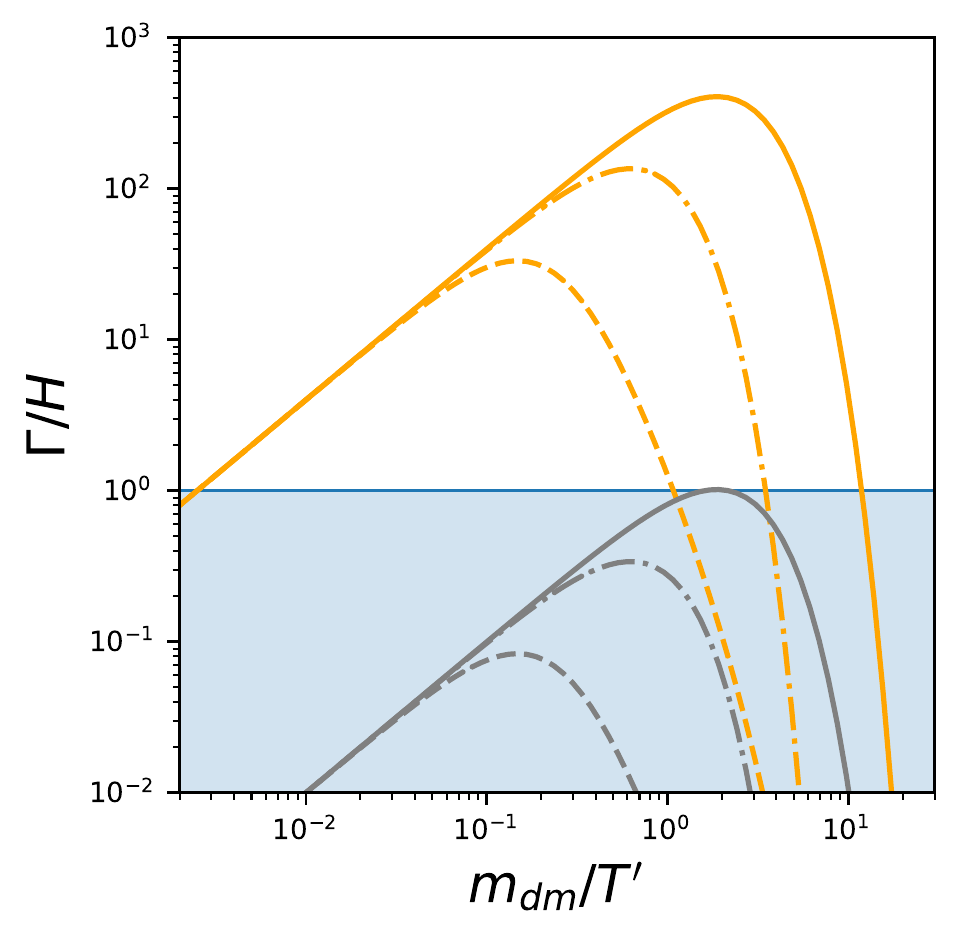}
		\caption{Schematic behaviour of interaction rates, $\Gamma$, divided by the expansion rate, $H \propto T^2$, as function of  $m_{\rm dm}/T'$. 
			DM particles in thermal equilibrium lie above $\Gamma/H =1$. 
			At high temperatures, $T' \gg m_{\rm dm}$, $\Gamma \sim T'$ in all cases. As the temperature of the HS drops, we consider three scenarios for DM decoupling, see text: 1) solid lines, the DM particles become non-relativistic, 2) dot-dashed lines, the DM {annihilates into} heavier particles of mass $m'$ (here chosen to be $m'/m_{\rm dm}=3$), and 3) dashed lines, the DM interacts through some heavy mediator and decouples while still relativistic. 
			We show these three cases for the same DM mass but for two choices of $\xi = T'/T$ (orange and grey lines, respectively). For the orange curves, the intersections with the horizontal line correspond to the temperatures $T'_{\rm eq}$ and $T'_{\rm dec}$. Between these two temperatures, the DM is in thermal equilibrium, $\Gamma > H$. The grey curves correspond to smaller $\xi$. In particular, the solid line is such that a DM  particle barely reaches thermal equilibrium around $T' \sim m_{\rm dm} \sim T_{\rm eq}= T'_{\rm dec}$.
		}
		\label{fig:rates}
	\end{figure}

	Consider again the case of SM neutrinos in the early universe. At high temperatures, larger than the electroweak scale, their interaction rate was $\Gamma \sim  \alpha_W^2 T$ while the expansion rate of the universe was $H \sim \sqrt{g_\ast} \,T^2/m_{\rm pl}$ \cite{Kolb:1990vq}. 
	Thus, SM neutrinos were in thermal equilibrium 
	at temperatures  $T \leq T_{\rm eq} \simeq \alpha_W^2\, m_{\rm pl}/ \sqrt{g_\ast} \sim 10^{14}$ GeV. 
	Translating this to relativistic particles in a HS interacting with a rate $\Gamma' \sim \alpha'^2  T'$, gives
	\begin{equation}
		\alpha'^2 T'_{\rm eq}\simeq  H(T_{\rm eq}) \Rightarrow T_{\rm eq} \simeq \alpha^{\prime 2}\xi_{\rm eq} \, m_{\rm pl}/ \sqrt{g_\ast}  \, ,
		\label{eq:TeqRel}
	\end{equation}
	where $\xi_{\rm eq} = (T^\prime/T)_{\rm eq}$ and $\alpha^\prime$ is a HS analogue of $\alpha_W$. So,  if $\xi < 1$ ($\xi >1$), particles in a HS entered  thermal equilibrium at a later (resp. earlier) time in the history of the universe than those in the VS. This is illustrated in Fig.~\ref{fig:rates}, where we plot $\Gamma/H$ as a function of 
	$x \equiv m_{\rm dm}/T'$, with thermal equilibrium corresponding to $\Gamma/H >1$. 
	
	The way DM decoupled is model-dependent. Here, we consider three generic possibilities. 
	The first is that the DM stayed in thermal equilibrium until it became non-relativistic and its abundance changed from $n_{\rm dm} \propto T'^3$ to being Maxwell-Boltzmann suppressed at $T' \lesssim m_{\rm dm}$, with $n_{\rm dm} \propto \exp(- m_{\rm dm}/T')$. This case is represented by solid lines in Fig. \ref{fig:rates}. 
	{The second and third possibilities are the two relativistic decoupling scenarios already discussed above, of annihilation cut off by a heavy mediator mass or by the mass of the particles that the DM scatters into.
		The corresponding rates are shown  in Fig.~\ref{fig:rates} as dashed lines and dot-dashed lines respectively. Their behaviour differs from the first case when $T'$ becomes smaller than the corresponding cut-off mass scale.}
	Other scenarios are possible, such as combinations of the above three cases. However, in what follows, and in particular in considering explicit models in section \ref{sec:models}, we will focus on the three simple possibilities considered here. 
	
	These three cases are depicted in Fig. \ref{fig:rates} assuming the same DM mass, but for two choices of $\xi$ (orange and grey lines). Candidates along the orange curves were in thermal equilibrium between $T^\prime_{\rm eq}$ and $T^\prime_{\rm dec}$. If the ratio $\xi$ increases (decreases), these curves move up (resp. down).  
	Thus, a generic feature which can be read from Fig. \ref{fig:rates} 
	is that the temperature range within which the DM was in chemical equilibrium shrinks as $\xi$ decreases. If $\xi$ is too small, all other things being kept the same, {then the} particle was never in equilibrium. 
	The absolute minimum value of $\xi$ for which there is thermalisation for a given DM model is therefore obtained by setting 
	$T'_{\rm eq} \approx T'_{\rm dec} (\approx m_{\rm dm})$, see
	the solid grey curve in Fig.~\ref{fig:rates}. This corresponds to the case of freeze-out of a mildly non-relativistic DM particle, thus it lies close to the relativistic floor. 
	
	To be specific, let us take $\Gamma \sim \alpha'^2 T'$ as above {when $T' \gg m_{\rm dm}$}. This leads to the $T'$ temperature range for thermal equilibrium\footnote{\label{fn:therm}This range assumes that the DM mass is the only relevant mass scale (the first of the three possibilities discussed above). If, instead, the DM is interacting with heavier particles of, say, mass $m'$, then the condition of thermalisation is more constraining as decoupling occurs for $T' \sim m'$. Thus $m_{\rm dm} \rightarrow m'$ in Eqs. (\ref{eq:range1},\ref{eq:range2},\ref{eq:therm}) and the thermalisation condition becomes independent of the DM mass for $m_{\rm dm} \lesssim m'$, 
		$$\xi_{\rm min} \approx \sqrt{g_\ast^{1/2} m'\over \alpha'^2 m_{\rm pl}}\,.$$ We will meet such a situation in section \ref{sec:models}. }
	\begin{equation}
		\label{eq:range1}
		T^\prime_{\rm dec} < T^\prime < T^\prime_{\rm eq} \Rightarrow m_{\rm dm} \lesssim T^\prime \lesssim \alpha^{\prime 2}\xi^2 {\, m_{\rm pl}\over \sqrt{g_\ast}}
	\end{equation}
	or equivalently,
	\begin{equation}
		\label{eq:range2}
		{m_{\rm dm}\over \xi} \lesssim T \lesssim \alpha^{\prime 2}\xi {\, m_{\rm pl}\over \sqrt{g_\ast}}
	\end{equation}
	This range shrinks to {zero as $\xi$ decreases} down to a minimum value found by taking $T_{\rm dec}' \approx T_{\rm eq}'$,  
	\begin{equation}
		\label{eq:therm}
		\xi_{\rm min} \approx \sqrt{ g_\ast^{1/2}  m_{\rm dm} \over \alpha'^2 m_{\rm pl}}\,.
	\end{equation}
	This condition of thermalisation of the HS
	crosses the relativistic floor, Eq.~(\ref{eq:floor}), 
	at 
	\begin{equation}
		\label{eq:limitTH}
		{m_{\rm dm} } \sim {\alpha^{\prime 6/5}}\; \mbox{\rm PeV}\,,
	\end{equation}
	corresponding to 
	\begin{equation}
		\xi_{\rm min} \sim 10^{-6}\,\alpha'^{-2/5} \,.
	\end{equation}
	As expected, $\xi$  {can be} lower if the DM and its companion particles have stronger interactions.  
	However, for fundamental particles, cross sections are constrained from above by unitarity \cite{Griest:1989wd}. Thus, even if  the DM abundance does not depend on the cross section in the case of relativistic decoupling, the DM mass along the relativistic  floor cannot be arbitrarily large due to the requirement of thermalisation.
	
	To determine the absolute limits on the DM mass along the relativistic floor, we consider the maximal, thermally averaged cross section allowed by unitarity. Assuming annihilation in $J=0$ state, it is given by \cite{Hambye:2020lvy}
	\begin{eqnarray}
		\label{eq:Unitarity1}
		\left\langle\sigma v\right\rangle &=& \frac{\pi}{4T'^{2}}\mathcal{I}_\epsilon\left(\frac{m_{\rm dm}}{T'}\right)
	\end{eqnarray}
	where $\mathcal{I}_\epsilon$ is
	\begin{eqnarray}
		\label{eq:Unitarity2}
		\mathcal{I}_\epsilon(x') &=& \frac{1}{N_\epsilon^{2}}\times\int_{4x'^{2}}^{\infty}\diff w\int_{\sqrt{w}}^{\infty}\diff k_{+}\int_{-k_{-,max}}^{k_{-,max}}\diff k_{-}\nonumber\\
		&&\times\left\lbrace\frac{\sqrt{w/(w-4x'^{2})}}{\left(e^{\frac{k_{+}+k_{-}}{2}}+\epsilon \right)\left(e^{\frac{k_{+}-k_{-}}{2}}+\epsilon \right)}\right\rbrace \label{eq:I_function}
	\end{eqnarray}
	with $N_\epsilon \equiv \int_{x'}^{\infty}\frac{\sqrt{k^{2}-x'^{2}}k\diff k}{e^{k}+\epsilon}$, where $\epsilon = 1$ for fermions, $-1$ for bosons and $0$ in the classical (Maxwell-Boltzmann) case, and $k_{-,max}=\sqrt{1-4x'^2/w}\sqrt{k_+^2-w}$. As we will consider regimes in which the DM is relativistic or mildly non-relativistic, it is important to keep track of quantum statistics effects. 
	In the relativistic limit, $T^\prime \gg m_{\rm dm}$, Eq. (\ref{eq:Unitarity1}) is
	\begin{equation}
		\left\langle\sigma v\right\rangle = {n_\epsilon\over T^{\prime 2}}
	\end{equation}
	with $n_{+1} = 5\pi/12$ and $n_{-1} = 15\pi/16$. For a classical Maxwell-Boltzmann distribution, (\ref{eq:Unitarity1}) takes the form 
	\begin{equation}
		\label{eq:unitaryXS}
		\left\langle\sigma v\right\rangle = {4 \pi \over m_{\rm dm}^{2}}\frac{K_2(2 x')}{K_2^2(x')}
	\end{equation}
	This gives $n_0 = \pi/2$ in the relativistic limit and $\langle 1/v \rangle = \sqrt{m_{\rm dm}/\pi T'}$ for $T'\ll m_{\rm dm}$, see Eq. (\ref{eq:unitarity}).
	
	Using Eqs.~(\ref{eq:Unitarity1})-(\ref{eq:Unitarity2}), we can find the lowest temperature $T'_{\rm dec} = T'_{\rm eq}$ at which a HS can be in thermal equilibrium, given the maximally allowed cross section of the DM. 
	This condition is depicted as  the orange region in Fig. \ref{fig:polygon}. It is set by simply requiring that 
	$$n_{\rm dm}\langle\sigma v\rangle \approx  H$$
	at $T'_{\rm dec} = T'_{\rm eq}$, which gives what we call the thermalisation condition,
	\begin{eqnarray}
		\xi_{\rm dec} &=& 1.2 \times 10^{-8}\,\left(\frac{m_{\rm dm}}{100 \text{ GeV}}\right)^{1/2}
		\left(\frac{\sqrt{g_{\ast,\rm dec}}}{g'}\right)^{1/2}\nonumber\\
		&&\hspace{2cm}\times\left(\frac{1}{x'_{\rm dec}\mathcal{I}(x'_{\rm dec})}\right)^{1/2}
		\label{eq:TH}
	\end{eqnarray}
	where $g_{\ast,\rm dec}$ counts the effective number of degrees of freedom contributing to the expansion rate at $T'_{\rm dec} = T'_{\rm eq}$.
	Modulo the dependence through $x'_{\rm dec} = m_{\rm dm}/T'_{\rm dec}$, we see that essentially $\xi_{\rm eq} \propto m_{\rm dm}^{1/2}$ as in (\ref{eq:therm}). As above, the thermalisation condition  crosses 
	the relativistic floor, Eq.~(\ref{eq:relxi}), at 
	\begin{equation}
		\label{eq:ximin1}
		\xi_{\rm min} \approx 1.4\times 10^{-5}\,,
	\end{equation}
	which {gives} the lowest possible temperature for a HS with thermal DM that decouples when (barely) relativistic. 
	This temperature corresponds a DM candidate of mass
	\begin{equation}
		\label{eq:massmax1}
		m_{\rm dm} \approx 30 \text{ PeV}\,.
	\end{equation}

	\subsubsection{Unitarity wall}
	\label{sec:theory_unitarity}
	If we depart from the point corresponding to (\ref{eq:massmax1}), going along the {orange} thermalisation condition, the HS temperature increases as $T'\propto m_{\rm dm}$, so that the possible DM candidates are less and less relativistic and we enter into a secluded regime of non-relativistic freeze-out \cite{Feng:2008mu,Chu:2011be}. 
	In this case, the relic abundance depends on the annihilation rate. 
	In the instantaneous freeze-out approximation, i.e.~fixing $x'_{\rm dec}=m_{\rm dm}/T'_{\rm dec}$ from the condition $\Gamma/H|_{T'=T'_{\rm dec}}=1$ and determining $Y_{\rm dm}$ by assuming that the yield after freeze-out is equal to $n_{\rm dm, eq}/s|_{T'=T'_{\rm dec}}$, the relic abundance obtained is\footnote{{It can be checked that the relic abundance obtained in this way differs from the one obtained from a proper integration of the Boltzmann equation
			by less than a factor $\sim 2$, regardless of the value of $\xi$ considered
			in the allowed domain of Fig.~\ref{fig:polygon}, and as long as $x'_{\rm dec} \gtrsim 3$. The agreement can be somewhat further improved 
			using the following  expression:
			$\Omega_{\rm dm}h^2=1.3 \cdot 10^8\, g_{\rm dm}\cdot(m_{\rm dm}/{\rm GeV})\,\left(\int_{x'_{\rm dec}}^\infty\langle\sigma v\rangle s/x'H\diff x'\right)^{-1}$.}} 
	\begin{eqnarray}
		\Omega _{\rm dm} h^{2} &=& 4.7\cdot10^8 \frac{g_{\rm dm} \,g_{\ast,\rm dec}^{1/2} \,x'_{\rm dec}\, \xi}{g_{\ast s,{\rm dec}}m_{\rm pl}\langle\sigma v\rangle\,{\rm GeV}}
		\label{Omeganonrelat}
	\end{eqnarray}
	with $x'_{\rm dec}$ given by\footnote{
		The power of $\xi$ in the logarithm which gives $x'_{\rm dec}$  can be obtained different from the value 2 in Eq. (\ref{eq:xp_dec}), if one uses another prescription for determining $x'_{\rm dec}$, see \cite{Feng:2009mn,Chu:2011be} for other prescriptions which lead to a value 
		5/2 or 3/2 in the $\xi_{\rm dec} <1$ case.}
	\begin{eqnarray}
		&&x'_{\rm dec} \simeq \ln\left[0.038\,\xi^2\left\langle\sigma v\right\rangle m_{\rm pl}m_{\rm dm}\left(\frac{g_{\rm dm}}{\sqrt{g_{\ast,\rm dec}}}\right)\right]\nonumber\\
		&&\hspace{0.5cm}+\frac{1}{2}\ln\ln\left[0.038\,\xi^2\left\langle\sigma v\right\rangle m_{\rm pl}m_{\rm dm}\left(\frac{g_{\rm dm}}{\sqrt{g_{\ast, {\rm dec}}}}\right)\right]\label{eq:xp_dec}\quad\quad
	\end{eqnarray}
	We first consider the case in which the expansion of the universe is dominated by the VS. This is natural for $\xi \ll 1$. In this case
	\begin{eqnarray}
		H(x') &=& 1.67\sqrt{g_{\ast}(T)}\frac{m_{\rm dm}^{2}}{m_{\rm pl}x'^{2}\xi^{2}}\, ,\\
		s(x') &=& \frac{2\pi^{2}}{45}g_{\ast s}(T)\frac{m_{\rm dm}^{3}}{x'^{3}\xi^{3}}\, .
	\end{eqnarray}
	Taking the maximal annihilation cross section allowed by unitarity, Eqs.~(\ref{eq:Unitarity1}) and (\ref{eq:Unitarity2}), gives an upper bound on $\xi_{\rm dec}$ as a function of $m_{\rm dm}$,
	\begin{eqnarray}
		\label{eq:newunitarity}
		\xi_{\rm dec}\leq \xi^{\rm unitar.}_{\rm dec} &\equiv& 1.18\cdot 10^{5}\;\left(\frac{100 \text{ GeV}}{m_{\rm dm}}\right)^{2}\nonumber\\
		&&\hspace{-1cm}\times \frac{g_{\ast s,\rm dec}}{\sqrt{g_{\ast,\rm dec}}}\; x'_{\rm dec}\;\mathcal{I}(x'_{\rm dec})\; 
		\label{eq:UB_VS}
	\end{eqnarray}
	This equation leads to the diagonal part of the {red} boundary in Fig.~\ref{fig:polygon}. For fixed $\xi_{\rm dec}$, {a value of the DM mass beyond this diagonal leads} to an excess of DM. Clearly, the relation (\ref{eq:newunitarity}) generalises the standard unitarity bound which is set assuming $T' = T$ \cite{Griest:1989wd}. 
	
	Let us make a few remarks. First we note that the  condition (\ref{eq:newunitarity}) scales as $\xi_{\rm dec} \propto 1/m_{\rm dm}^2$, so heavier DM candidates are possible for $T' < T$. Second, this expression, which is obtained for the unitarity limited cross section, implies that in general $\xi_{\rm dec} \propto \langle \sigma v \rangle$. Indeed, for fixed DM mass, decreasing the cross-section by decreasing the coupling leads to a increase of the DM abundance (as for standard non-relativistic FO at $T' =T$), which is compensated by a decrease of the HS temperature.  
	Lastly, we see from Eq. (\ref{eq:xp_dec}) that $x'_{\rm dec}$ decreases as the DM mass increases when $\langle \sigma v \rangle \propto m_{\rm dm}^{-2}$. While  $x'_{\rm dec} = {\cal O}(20)$ for $T' \sim T$, the DM is less non-relativistic at decoupling if $T' \ll T$, {as expected since in this case the DM particles are less numerous already to start with and thus must be less Boltzmann suppressed to account for the relic density.}
	
	The crossing of the unitarity constraint (\ref{eq:newunitarity}) and the condition for thermalisation (\ref{eq:TH}) gives
	the heaviest possible thermal DM candidate (denoted by the little red dot in Fig.~\ref{fig:polygon}), which has mass 
	\begin{equation}
		\label{eq:massmax2}
		m_{\rm dm,max} \approx 52 \, \mbox{\rm PeV}
	\end{equation}
	when DM is a Dirac fermion. 
	This is slightly heavier than (\ref{eq:massmax1}) and corresponds to a temperature ratio, 
	\begin{equation}
		\label{eq:ximin2}
		\xi \approx 6.9 \cdot 10^{-5}
	\end{equation} 
	to be compared with (\ref{eq:ximin1}). It may be worth mentioning that taking freeze-out at $T'_{\rm dec} \sim m_{\rm dm,max}$ implies that the DM decoupled around $T \sim 10^{11}$ GeV. 
	
	So far, we have assumed that the expansion of the universe was dominated by the VS. If there are many degrees of freedom in the HS or if $\xi \gtrsim 1$, the  entropy and energy densities of the HS can be dominant at the time of the decoupling of the DM.  
	In this case, we can approximate the expansion rate and entropy by
	\begin{eqnarray}
		H(x') &\approx& 1.67\sqrt{g'_\ast(T')}\frac{m_{\rm dm}'^{2}}{m_{\rm pl}x'^{2}},\label{Hforlargexi}\\
		s(x') &\approx& \frac{2\pi^{2}}{45}g'_{\ast s}(T')\frac{m_{\rm dm}^{3}}{x'^{3}} 
		\label{sforlargexi}
	\end{eqnarray}
	by  neglecting the VS contributions, where $g'_\ast$ is the effective number of HS relativistic degrees of freedom. In this approximation, the DM abundance no longer depends on $\xi$. Put simply, the VS, and its temperature, are irrelevant at the time of DM decoupling. Thus, because the annihilation cross section is set by unitarity, there is a unique value for the maximum DM mass as long as the HS dominates the universe.

	In this scenario it is crucial that after DM freeze-out, most of the entropy and energy of the HS, which is shared by the DM companions, is transferred to the VS. There are two main possibilities, depending on whether or not the entropy is conserved in this transfer:\newline
	\noindent - If the transfer occurs while the companions are in thermal equilibrium with the VS, entropy is conserved and Eq.~(\ref{Omeganonrelat}) together with Eqs.~(\ref{Hforlargexi}) and (\ref{sforlargexi}) apply. 
	This gives the upper bound on $m_{\rm dm}$ corresponding to the vertical part of the red exclusion region in Fig.~\ref{fig:polygon}, 
	\begin{eqnarray}
		m_{\rm dm} &\approx& 35\, \text{TeV}\,g_{\rm dm}^{1/4}\,\left(x'_{\rm dec}\mathcal{I}(x'_{\rm dec}) \right)^{1/2}\label{eq:UB_HS}
	\end{eqnarray}
	To draw this limit, we considered a minimal scenario, with ${\cal O}(1)$ companion particles on top of the DM and only the known, SM degrees of freedom in the VS. 
	It intersects the curve (\ref{eq:newunitarity}) at roughly $\xi  \approx \sqrt[4]{g_\ast/g'_\ast} \approx 3$. 
	If there were more particles in the VS than the ${\cal O}(100)$ of the SM, the line would shift towards the left but, as it scales as the fourth root of the number of particles, their effect would be  mild.\newline
	\noindent - If transfer happens {through a slow, out-of-equilibrium decay of the companion particles, entropy is produced.}
	Energy conservation leads to heating of the VS \cite{Scherrer:1984fd} and dilution of the DM abundance. Such a scenario has been considered in \cite{Berlin:2016gtr}. {The effect of entropy production is to slightly shift the vertical line in Fig.~\ref{fig:polygon} to the right. The shift can be approximated by multiplying (\ref{eq:UB_HS}) by a factor of}
	\begin{eqnarray}
		f\simeq\left(\frac{g_{\ast s}^{4}}{g'_{\ast}g_{\ast}^{3}}\right)^{1/8}\label{eq:UB_HS-rho} \geq 1
	\end{eqnarray}   
	{obtained by imposing that $\rho_{\rm tot}=\rho_{VS}+\rho_{HS}$ is conserved and then extracting the evolution of the temperature ratio due to this conservation. As (\ref{eq:UB_HS}) is {derived} assuming entropy conservation, the factor (\ref{eq:UB_HS-rho}) divides by the contribution coming from entropy conservation ($g_{\ast s}$, as defined in Eq. (\ref{eq:relxi})) and multiplies the contribution from energy conservation ($g_{\ast}$). This factor lies between $f\sim 1$ and $f\sim 1.6$ depending on the VS temperature when the entropy is injected ($T<10$ keV and $T>$ TeV respectively) for $g_\ast' = \mathcal{O}(1)$}.
	{We note at this stage also that the DM abundance from relativistic freeze-out given by Eq. \eqref{eq:relxi} also assumed entropy conservation. 
		An out-of-equilibrium processes resulting in entropy production would cause the RHS of Eq. \eqref{eq:relxi} to be a factor of $(g'_{\ast s,\rm dec}/g'_{\ast s,\rm in})^{1/4}$ smaller. 
		Since $g'_{\ast s,\rm dec} \leq g'_{\ast s,\rm in}$, this could slightly raise the relativistic floor.}

	\subsection{Observational constraints}
	\label{sec:exp}
	
	In this section, we consider two observations that set important restrictions on the domain of thermal DM candidates. These are depicted in Fig.~\ref{fig:polygon}, see the green ($N_{\rm eff}$ ceiling) and purple (free streaming) regions. Our aim, as in the previous section, is to be as model-independent as possible.
	\subsubsection{N$_{\rm eff}$ ceiling}
	\label{sec:neff}
	
	We know that the  universe was dominated by radiation at the time of big bang nucleosynthesis (BBN) {and until about the recombination epoch.}. 
	A crude constraint is set by considering the possible contribution of {the HS particles}
	to the number of relativistic degrees of freedom at the time of BBN {and recombination}. This is expressed in terms of $ N_\text{eff}$, the effective number of neutrinos. Too large a value of $\Delta N_\text{eff}$ at temperatures around $T\sim \mathcal{O}(1)$ MeV will increase the Hubble rate and thereby impact the abundances of light nuclei, which are rather well {measured. 
		The latest} CMB measurement by Planck \cite{Aghanim:2018eyx} gives $N_\text{eff} = 2.99 \pm 0.17$ (the results from BBN are similar \cite{Fields:2019pfx}), which is to be compared with the SM prediction, $N_\text{eff} = 3.04$, see e.g. \cite{Bennett:2019ewm}. Hence, we impose the constraint that $\Delta N_\text{eff} \leq 0.29$ at $2\sigma$. 
	We will distinguish two contributions to $\Delta N_\text{eff}$, one from the DM degrees of freedom and one from other HS particles. The latter is much more model-dependent than the former.
	
	We first look at the contribution of the DM degrees of freedom. 
	The shift in $N_\text{eff}$ due to the dark matter at $T = 1$ MeV can be computed by taking the ratio of the DM and neutrino energy densities. 
	Using the fermionic (+) or bosonic (-) equilibrium number densities, we can {express $\Delta N_\text{eff}$ as
		\begin{eqnarray}
			\Delta N_\text{eff} &=& {8\over 7} \left({11\over 4}\right)^{4/3} \frac{\rho_{\rm dm}}{\rho_{\gamma}}= \frac{60 g_{\rm dm} \xi^4}{7\pi^4} \left( \frac{11}{4} \right)^{4/3} \notag \\
			&\times &\int_{\frac{m_{\rm dm}}{T'_{\rm BBN}}}^\infty \!\!dz\; \frac{z^2 \sqrt{z^2 - (m_{\rm dm}/T'_{\rm BBN})^2}}{e^{z} \pm 1} \, ,
			\label{eq:neff}
		\end{eqnarray}
		where $z = E_{\rm dm}/T'$ with $T'_{\rm BBN}=\xi\,T_{\rm BBN}$}
	Any other relativistic degrees of freedom in the HS will give an additional contribution to $\Delta N_\text{eff}$, so we are being very conservative here. 
	The constraint it leads to is depicted by the green region of Fig. \ref{fig:polygon}. 
	Its two key features may easily be understood. 
	Firstly, it is clear from Eq. \eqref{eq:neff} that the contribution of a DM particle to $\Delta N_\text{eff}$ is only sizeable when it is relativistic at the time of BBN, $T_{\rm BBN} \sim 1 \text{ MeV}$, i.e. for $\xi \gtrsim m_\text{dm}/\text{MeV}$. 
	{This basic condition sets the diagonal part of the green boundary for $m_{\rm dm} \gtrsim 10$ MeV. }
	Next, suppose that DM is indeed relativistic at BBN, i.e. $m_\text{dm} \ll  T'_{\rm BBN}$. For fermionic DM, we have $\Delta N_\text{eff} \simeq g_{\rm dm} (11/4)^{4/3} \xi^4/2$, meaning that a Dirac fermion ($g_{\rm dm}=4$) counts as two families of neutrinos, while a Majorana DM particle ($g_{\rm dm}=2$) counts as one. For bosonic DM instead, $\Delta N_\text{eff} \simeq  4 g_{\rm dm} (11/4)^{4/3} \xi^4/7$ (with $g_{\rm dm}=1$ for a real scalar). Collectively, we write this as $\Delta N_\text{eff} = 4 g_{\rm dm,eff} (11/4)^{4/3} \xi^4 /7$ with $g_{\rm dm,eff} = \sum_{B} g_{\rm dm,B} + 7/8 \sum_{F} g_{\rm dm,F}$. The bound from Planck then corresponds to
	\begin{equation}
		\label{eq:planck1}
		\xi \lesssim {0.60}/(g_{\rm dm,eff})^{1/4} \, , \qquad m_{\rm dm} \ll \xi\, T_{\rm BBN} \, ,
	\end{equation}
	which excludes any value of $\xi$ larger than $\sim 1$ when $m_\text{dm} \ll  T'_{\rm BBN}$. As soon as $T'$ is below $T$, however, the $\xi^4$ suppression allows
	a large number of relativistic degrees of freedom \cite{Feng:2008mu}.
	Notably, the constraint \eqref{eq:planck1} is independent of the DM mass. This limit corresponds to the horizontal part of the green region. To draw it, we assumed $g_{\rm dm,eff} = {3.5}$ (i.e. a Dirac fermion), but taking instead $g_{\rm dm,eff} = 10$, for instance, barely changes the figure.
	
	Next, we discuss the possible additional contribution of companions in the HS. This is much more model-dependent as it is related to the number of particles which are left after DM freeze-out and how their number changes afterwards. Consider first the typical situation in which DM annihilates into lighter particles which decouple relativistically.
	{In this case, if they disappear (say, by decaying to SM particles) before the BBN epoch they are harmless for BBN. 
		However, since these particles will not disappear 
		{until they are non-relativistic}, this requires their mass  
		to be larger than $T'_{\rm BBN}$.}
	In this case, if the mass of the companion particle is about the DM mass the constraint remains as given by the green area in Fig.~\ref{fig:polygon}. If the companion particle is lighter than the DM by a fixed ratio, the diagonal part instead moves to the right by a factor of $m_{\rm dm}/m_{\rm comp}$. Alternatively, for a fixed value of $m_{\rm comp}$, one gets an horizontal line, i.e.~a fixed upper bound on the temperature ratio: $\xi < m_{\rm comp}/T_{\rm BBN}$.
	In the extreme case, where the companion is much lighter than the DM particle and/or it decays after BBN,  
	the bound \eqref{eq:planck1} becomes
	\begin{equation}
		\label{eq:planck2}
		\xi \lesssim 0.60/(g_{\rm dm,light})^{1/4} 
	\end{equation}
	where $g_{\rm dm, light}$ now includes all HS degrees of freedom that are still abundant at the BBN time (see also \cite{Feng:2008mu}).
	This corresponds to a horizontal exclusion line which for low DM mass is somewhat below the horizontal boundary of the green region in Fig. \ref{fig:polygon} (due to a larger number of degrees of freedom contributing in Eq.~(\ref{eq:planck2}) than in Eq.~(\ref{eq:planck1})),
	and extends horizontally for larger DM masses (with a somewhat small step up at $m_{\rm dm}\sim 1$~MeV, since above this value DM ceases to contribute). Thus, to first approximation, (\ref{eq:planck2}) leads to an extension of the horizontal green exclusion region towards larger DM mass, which is depicted by the dashed green line in Fig. \ref{fig:polygon}. 
	To 
	summarise, the region allowed by BBN can, if there are no HS degrees of freedom left at BBN time, extend to the white domain depicted in Fig. \ref{fig:polygon}. If this is not the case, values of $\xi$ larger than $\sim 1$ are excluded.
	
	The above discussion was deliberately general. For specific models, the constraints can be expected to be stronger than the generic BBN bound we discussed here. The most delicate situation is when the DM and/or its companion particles annihilate or decay at the time of BBN, as the production of light elements can be affected by changes to the expansion rate and, more importantly, by energy transfer into the VS, see e.g. \cite{Berger:2016vxi,Hufnagel:2017dgo,Hufnagel:2018bjp}. Determining this, however, necessarily demands specifying a HS scenario.

	\subsubsection{Free-streaming}
	\label{sec:streaming}
	Thermal DM cannot be too light, otherwise it remains relativistic for too long and does not permit the formation of large-scale structures that we observe. 
	For a thermal relic with $\xi = 1$, the strongest bound is $m_{\rm dm} > 5.3$ keV \cite{Irsic:2017ixq}, obtained from Lyman-$\alpha$ forest data. 
	This bound can be generalised to account for different values of $T'/T$ by converting it to a limit on the DM free-streaming horizon, the average {distance a 
		DM particle} travels after production. 
	
	The average momentum of a population of particles in thermal equilibrium with temperature $T'$ is
	\begin{align}
		\langle p \rangle \equiv \frac{\int d^3 p f(p) p}{\int d^3p f(p)} = T' \times \begin{cases}
			3.15 \text{ for a fermion,} \\
			2.70 \text{ for a boson.}
		\end{cases}
	\end{align}
	This average persists after the population goes out of thermal equilibrium. 
	The particle species can then be said to be non-relativistic when $\langle p \rangle \leq m$, i.e. below $T'_{\rm nr} = m_{\rm dm}/3.15$ for a fermion, 
	with the corresponding time being $t_{\rm nr} = 1/[2H(T'_{\rm nr}/\xi)]$. 
	If $t_{\rm nr} < t_{\rm eq} = 1.9 \times 10^{11}$s, the time of matter-radiation equality, then the free-streaming horizon can be estimated as \cite{Kolb:1990vq}
	\begin{align}
		\lambda_{\rm FS} &= \int_{t_{\rm dec}}^{t_0} \frac{\langle v(t) \rangle}{a(t)} dt \label{eq:fs} \\
		&\simeq \frac{\sqrt{t_{\rm eq} t_{\rm nr}}}{a_{\rm eq}} \left( 5 + \log \frac{t_{\rm eq}}{t_{\rm nr}} \right) \left( \frac{g_{\ast s,0} + \xi^3 g_{\ast s,0}'}{g_{\ast s,{\rm dec}} + \xi^3 g_{\ast s,{\rm dec}}'} \right)^{1/3} ~, \notag
	\end{align}
	where $a_{\rm eq} = 8.3 \times 10^{-5}$ is the scale factor at $t_{\rm eq}$. 
	On the other hand, if $t_{\rm nr} > t_{\rm eq}$, we have
	\begin{align}
		\lambda_{\rm FS} &\simeq \left[ \frac{6(t_{\rm eq}^2 t_{\rm nr})^{1/3} - t_{\rm eq}}{a_{\rm eq}} \right] \left( \frac{g_{\ast s,0} + \xi^3 g_{\ast s,0}'}{g_{\ast s,{\rm dec}} + \xi^3 g_{\ast s,{\rm dec}}'} \right)^{1/3} ~,\label{eq:fs2}
	\end{align}
	since for $t> t_{\rm  nr}$, $\langle v (t) \rangle \simeq a_{\rm nr}/a \simeq (t_{\rm nr}/t)^{2/3}$. 
	Note that Eqs. \eqref{eq:fs} and \eqref{eq:fs2} coincide when $t_{\rm nr} = t_{\rm eq}$.

	An early decoupled fermionic thermal relic of mass $5.3$ keV has a free-streaming horizon $\lambda_{FS} \simeq 0.066$ Mpc. 
	Imposing this upper limit on the free-streaming horizon leads to the bound given by the purple region in Fig.~\ref{fig:polygon}. 
	The behaviour is quite different depending on whether $\xi$ is smaller or larger than 1. 
	When $T' \ll T$, then $t_{\rm nr} \propto \xi^2/m_{\rm dm}^2$, hence $\lambda_{\rm FS} \propto \xi/m_{\rm dm}$, up to the logarithm in Eq. \eqref{eq:fs} (see also \cite{Hambye:2020lvy}) . 
	Conversely, when $T' \gg T$, the hidden sector dominates, with 
	$H(T'_{nr}) \sim m_{\rm dm}^2/m_{\rm pl}$. 
	Then if $t_{\rm nr} < t_{\rm eq}$, as is the case for $m_{\rm dm} \gtrsim 10$ eV, we have 
	$\lambda_{\rm FS} \propto 1/(\xi m_{\rm dm})$, neglecting the logarithm. As a result of all that, the absolute minimum value of $m_{\rm dm}$ lies at the intersection of the free-streaming constraint and the relativistic floor, which {is 
		at $m_{\rm dm} \simeq 1.0$~keV and $\xi \simeq 0.16$.
		Thus we obtain
		\begin{equation}
			m_{\rm dm} \gtrsim 1.0~\,\hbox{keV}\,,
		\end{equation}
		see also \cite{Hambye:2020lvy}.}

	\subsection{Domain of thermal DM candidates}
	
	\begin{figure}
		\centering
		\includegraphics[height=7.5cm]{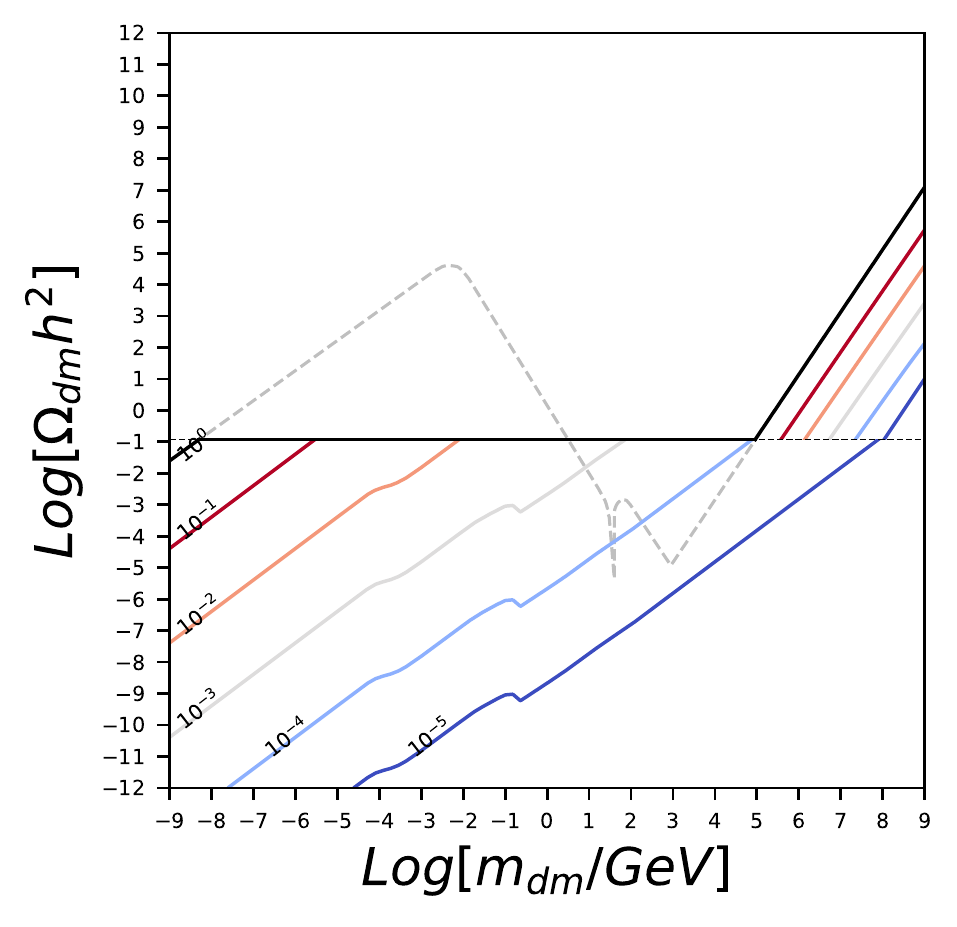}
		\caption{DM density 
			versus DM mass for various temperature ratios, $T'/T$. The solid black line is for $T' =T$ (the dashed line corresponds to the case of heavy neutrino DM, see Fig.~\ref{fig:neutrinos}).
		}
		\label{fig:polygon2}
	\end{figure}

	We summarise here the constraints discussed in sections \ref{sec:theory} (theory) and \ref{sec:exp} (observations). 
	Fig. \ref{fig:polygon} depicts the bounds obtained from the $\xi = T'/T$ relativistic floor (Eq. \ref{eq:floor}, boundary of blue region), the thermalisation constraint (imposing unitarity,  Eq. \ref{eq:TH}, orange) and the unitarity wall (Eqs. \ref{eq:UB_VS} and \ref{eq:UB_HS}, red). 
	Along the $\xi = 1$ horizontal dashed black line lies the domain  of DM candidates that were in equilibrium with the SM bath when they decoupled, Eq. (\ref{eq:1D}). 
	The observational constraints come from the $N_{\rm eff}$ ceiling  (green region, Eq. \ref{eq:neff}, assuming the products of DM annihilation are gone by the time of BBN) and from free-streaming (purple region, Eqs. \ref{eq:fs} and \ref{eq:fs2}). 
	All these constraints define the possible domain of thermal DM, as given by the white region. 
	In specific cases, this region can be further reduced. For instance, the dashed green line gives the approximate (somewhat model-dependent) upper bound on $\xi$ which holds when the (or some of the) companion(s) of the DM are still around during BBN. 
	The {vertical part of the red  region} corresponds to Eq. \eqref{eq:UB_HS} rather than Eq. \eqref{eq:UB_HS-rho} (i.e. for $\xi>1$, conservation of entropy rather than of energy when the HS particles disappear into SM particles later on) {but the difference is small}.

	In the blue region, bounded by the relativistic floor, the thermal DM particles would be under-abundant, while in the red region, bounded by the unitarity wall, they would be over-abundant. Only in between can viable thermal DM candidates exist. To make this clear, consider  Fig.~\ref{fig:polygon2}, in which we show the DM parameter density against the DM mass for different choices of $\xi = T'/T$. The horizontal dashed black line corresponds to the observed density, $\Omega_{\rm dm} h^2 \approx 0.12$. 
	The solid curves below this line correspond to particles that decoupled when relativistic for different choices of $\xi$ (from $\xi =1$ to $\xi = 10^{-5}$). {Thus they give the maximum value of $\Omega_{\rm dm}$ one can obtain for the values of  $m_{\rm dm}$ and $\xi$ considered.} The intersections of these curves with the horizontal line define the relativistic floor. The solid curves above the dashed line correspond to particles that decoupled when non-relativistic, assuming the unitarity bound for their annihilation cross {section (thus giving the minimum value of $\Omega_{\rm dm}$ one can obtain for the values of  $m_{\rm dm}$ and $\xi$ considered)}. The intersections of these curves with the dashed line define the unitarity wall. For fixed $\xi$, the line of given $\xi$ that overlaps the horizontal dashed line defines the corresponding thermal DM mass range within which it may be possible to have the observed relic density; this is illustrated by the horizontal black solid line in the case of $\xi=1$.  As explained above, and as can also be seen on this figure, as $\xi$ decreases the DM mass range shrinks and also shifts toward higher DM masses. Around $\xi \sim 10^{-5}$, the non-relativistic floor and unitarity wall merge (see blue curves), corresponding to DM candidates around the PeV scale. All together, they form the domain of thermal DM candidates, as depicted in the plane $\xi$ vs. $m_{\rm dm}$ in Fig. \ref{fig:polygon}.
	
	How specific models fit into this picture is the subject of the next section but, as an illustration, in Fig.~\ref{fig:polygon2} we reproduce the case of a Dirac neutrino for $\xi=1$, see the grey line and Fig. \ref{fig:neutrinos}. As discussed in section \ref{sec:1D}, this theory has three thermal DM candidates. The one that lies between the floor and the wall (the Lee-Weinberg candidate around $m_{\rm dm} \sim $ GeV) will move around if, forgetting for the sake of the argument the relation with electroweak interactions, we change its interactions, keeping $\xi=1$. For instance, increasing $\alpha_W$ will make it moves toward lighter DM while increasing the mass of the gauge boson(s) moves it toward heavier DM, etc. One may further convince oneself that, playing with the parameters of this model, the horizontal black solid interval may be filled by thermal candidates.
	
	\section{Explicit models}
	\label{sec:models}
	In this section, we  illustrate how some concrete models of hidden DM fit in the domain of thermal DM candidates of Fig.~\ref{fig:polygon}. We will do so using three simplified DM scenarios. In the first one, fermionic DM particles annihilate into a pair of vector bosons. 
	We dub this the t-channel scenario {after the topology of the tree-level annihilation process}. In the second scenario, the DM annihilates into HS fermions through an s-channel {process}.  Finally, we consider {scalar} DM annihilation into HS scalars via a contact interaction. We will skim over the fate of the DM companions, except in the last scenario.

	\subsection{Scenario 1 : t-channel}
	For this first model, the DM consists of a Dirac fermion, $\chi$, charged under a local $U(1)'$. The dark photon has mass $m_{\gamma'}$; the origin of its mass is not important here, but $m_{\gamma'}$ can be smaller or larger than the mass of the $\chi$, called $m_{\rm dm}$ as above. 
	We consider separately the cases $\xi < 1$ and $\xi > 1$. 
	\begin{figure}
		\centering
		\includegraphics[height=7.5cm]{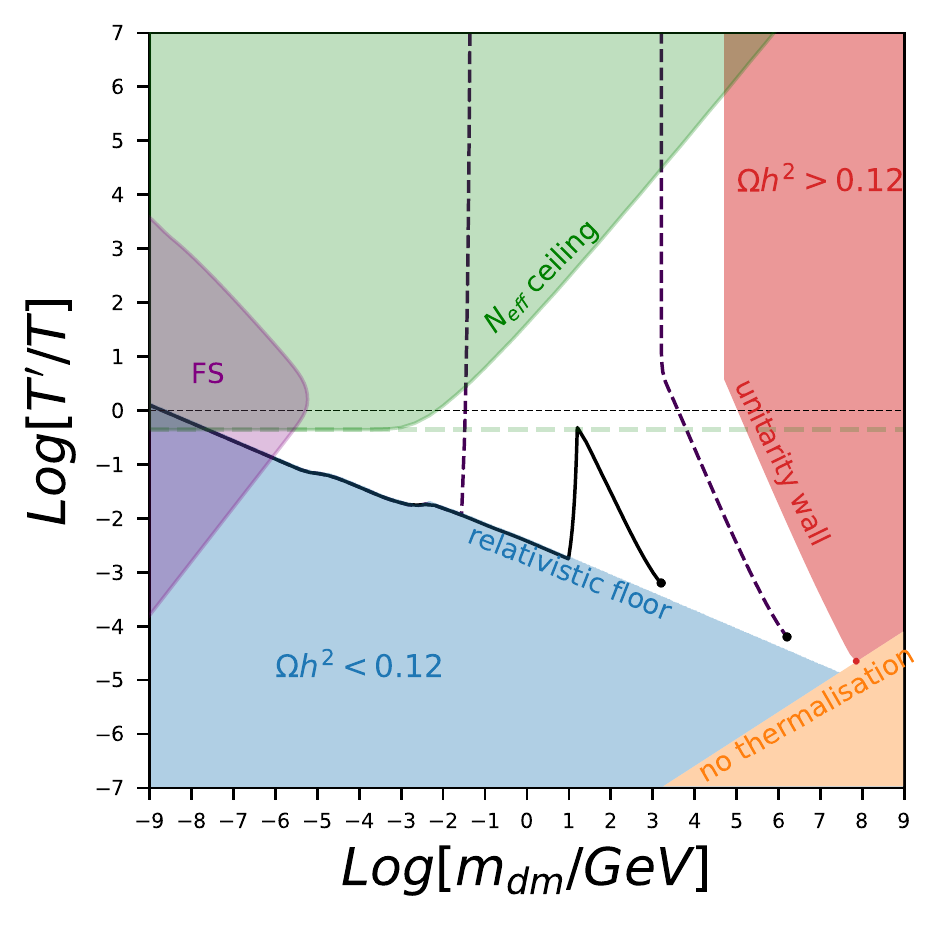}
		\caption{Parameter space for DM freeze-out driven by DM annihilating into two dark photons in the t-channel. The parameters are $\alpha'=3\cdot 10^{-4}$, $m_{\gamma'}=10$ GeV for the solid curve and $\alpha'=0.1$, $m_{\gamma'}=30$ MeV for the dashed curve.}\label{fig:polygon_t_ch_noDP}
	\end{figure}

\begin{figure}
		\centering
		\includegraphics[height=7.5cm]{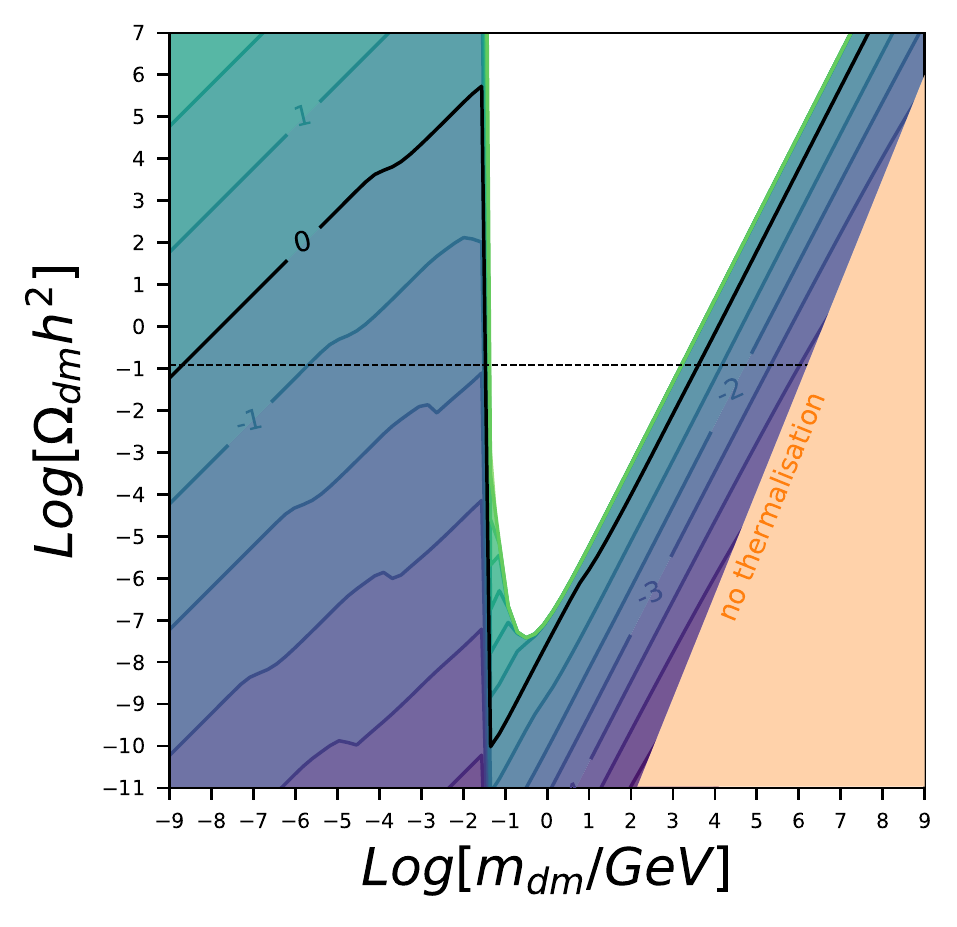}
		\caption{Parameter space for DM freeze-out driven by DM annihilating into two dark photons in the t-channel
			for $\alpha'=0.1$ and $m_{\gamma'}=30$ MeV. The contours correspond to different values of $\log(\xi)$.}\label{fig:polygon_t_ch}
	\end{figure}

	1) If $\xi < 1$, the DM and the dark photons have little impact on the expansion rate.  For fixed DM abundance, HS coupling $\alpha'$ and dark photon mass, the temperature ratio $\xi$ can be expressed as a function of the DM mass, $\xi =\xi(m_{\rm dm})$. 
	An example of such a relation is the black solid curve in 
	Fig.~\ref{fig:polygon_t_ch_noDP}. 
	Its salient features are the following:
	
	- For $m_{\rm dm} < m_{\gamma'}$, the DM particles can efficiently annihilate into the $\gamma'$ only as long as $T' \gtrsim m_{\gamma'}$. The annihilation becomes Boltzmann suppressed for  $T' \lesssim m_{\gamma'}$, so the DM particles decouple relativistically around 
	$T'\sim m_{\gamma'} > m_{\rm dm}$. 
	Thus, the only way to account for the relic density is if DM lies on the relativistic floor, see Fig.~\ref{fig:polygon_t_ch_noDP}.
	This can further be seen in Fig.~\ref{fig:polygon_t_ch} {(a model-specific version of Fig. \ref{fig:polygon2})}, which displays the DM abundance as a function of the DM mass for several different values of $\xi$. For $m_{\rm dm} < m_{\gamma'}$, the DM density parameter is $\propto m_{\rm dm}$, a signature of relativistic decoupling, so the candidates lie along the relativistic floor. 
	
	- At the threshold $m_{\rm dm} \sim m_{\gamma'}$, the annihilation rate into dark photons, which  was 
	{strongly Boltzmann suppressed at $T'\simeq m_{\rm dm}$ for $m_{\rm dm} < m_{\gamma'}$ (proportionally to $e^{-2m_{\gamma'}/m_{\rm dm}}$), 
		increases very quickly.} 
	The {resulting} efficient annihilation into dark photons leads to a drop in the abundance for fixed $\xi$, as can be also seen in Fig.~\ref{fig:polygon_t_ch}. Thus, to account for the relic density one needs a larger value of $\xi$, 
	as displayed in Fig.~\ref{fig:polygon_t_ch_noDP}. This marks a sharp transition from the relativistic freeze-out regime  
	to the non-relativistic freeze-out {one, i.e.~from the relativistic floor value of $\xi$ to the value of $\xi$ needed when the annihilation rate is no longer Boltzmann suppressed by the higher value of $m_{\gamma'}$.} 
	Note that such a threshold feature is not automatically present. Indeed, if the coupling decreases, it may no longer be possible to deplete the abundance through non-relativistic freeze-out. In this case, the curves of the DM candidates lie along the relativistic floor {up to a value of $m_{\rm dm}$ (below $m_{\gamma'}$), above which 
		DM can no longer be in thermal equilibrium, see below.} 
	
	- Well above the threshold, $m_{\rm dm} > m_{\gamma'}$, the DM lies in the secluded, non-relativistic freeze-out regime, $\Omega_{\rm dm}
	\propto m^2_{\rm dm}/\alpha'^2$, see Fig. \ref{fig:polygon_t_ch}, and the relationship between $\xi$ and $m_{\rm dm}$ evolves similarly to Eq.~\eqref{eq:newunitarity}, with $\xi_{\rm dec} \propto \alpha'^2/m_{\rm dm}^{2}$. The dependence on $\alpha'$ stems from the discussion below Eq.~\eqref{eq:newunitarity}, so the impact of changing $\alpha'$ is manifest. For instance, for $\xi < 1$, compare this solid line with the part of the black dashed line which lies in the $\xi <1$ region: it is obtained for a larger value of $\alpha'$ and  consequently lies at larger values of $m_{\rm dm}$. 
	
	- Still along these solid and dashed curves, as the DM mass increases and so $\xi$ decreases, FO occurs for larger $T'_{\rm dec}$ and smaller $x'_{\rm dec} = m_{\rm dm}/T'_{\rm dec}$, see Eq. (\ref{eq:xp_dec}). Eventually, FO takes place when the DM is only mildly non-relativistic, $m_{\rm dm} \gtrsim T'$. This mildly non-relativistic FO is the reason for the upturn of the solid and dashed curves as they go close to the relativistic floor. However, around such masses, thermalisation of the HS is no longer guaranteed, see \ref{sec:therm} and in particular Fig.~\ref{fig:rates}. Lack of thermal equilibrium implies that the curves stop at some DM mass, see the black dots on the solid and dashed black curves and Fig.~6 of \cite{Hambye_2020}. This dot corresponds to the intersection of the solid curve with the thermalisation diagonal which can be drawn for the given value of $\alpha'$, Eq.~(\ref{eq:therm}) (this is not drawn in Fig. \ref{fig:polygon_t_ch} but is simply parallel to the orange thermalisation line, given by Eq.~(\ref{eq:TH})).
	
	- The condition for thermal equilibrium scales like $\xi_{\rm min} \propto \sqrt{m_{\rm dm}/\alpha'^2}$, see Eq. (\ref{eq:limitTH}). {Thus,} as $\alpha'$ changes, the endpoint runs parallel and close to the relativistic floor. It ends in the lower right corner of the domain (the little red dot) for maximal coupling, where the unitary wall and the thermalisation condition meet. If, on the other hand, we decrease $\alpha'$, the thermalisation endpoint moves up, eventually reaching $m_{\rm dm} \approx m_{\gamma'}$. At this point, the condition for thermal equilibrium becomes $\xi_{\rm min} \propto \sqrt{m_{\gamma'}/\alpha'^2}$ (see footnote \ref{fn:therm}) so that for smaller $\alpha'$ the thermalisation endpoints run along the relativistic floor. 
	
	- We pointed out in section \ref{sec:1D} that, {varying the DM mass and} all other things being kept constant (here $\xi$, $m_{\gamma'}$ and $\alpha'$), we can in general expect to have an odd number of DM candidates, except at some fined tuned points (corresponding here to the regime $m_{\rm dm} \approx m_{\gamma'}$).
	{This can be seen from the fact that most contours in Fig. \ref{fig:polygon_t_ch} cross the $\Omega_{\rm dm}h^2 = 0.12$ line either once or thrice.}
	
	2) So far, we assumed that the model parameters are such that the DM candidates lie in the region $\xi < 1$.  
	Large values $\alpha'$ brings in DM candidates for which $\xi > 1$, see the dashed black curve in Fig.~\ref{fig:polygon_t_ch_noDP}. 
	We recognise on this curve several patterns that are similar to the case of smaller $\alpha'$. 
	Starting from the left of the figure, with low mass DM, the abundance along the solid line is set by relativistic decoupling until $m_{\rm dm} \approx m_{\gamma'}$ , at which point the annihilation channel opens up, leading to a Boltzmann suppression of the relic density that must be compensated by a sharp rise of $\xi$. 
	Also, as for smaller values of $\alpha'$, at large values of $m_{\rm dm}$ the candidates follow a diagonal line, $\xi_{\rm dec} \propto \alpha'^2/m_{\rm dm}^{2}$ (parallel to the unitarity wall), corresponding to a non-relativistic secluded freeze-out regime. 
	The line is closer of the unitarity wall than the solid line because  $\alpha'$ is larger. 
	
	{There is nevertheless a clear} difference between the solid and dashed lines for intermediate values of $m_{\rm dm}$ corresponding to the $\xi >1$ region. 
	For large $\alpha'$, the line extends itself to this region because as soon as the annihilation channel opens up at $m_{\rm dm} \approx m_{\gamma'}$, the large annihilation cross section leads to a large Boltzmann suppression factor which can only be compensated by a value of $\xi \gg 1$. In this case, however, the DM particle and the dark photons dominate the expansion rate, so the DM abundance no longer depends on the temperature of the visible sector, see section \ref{sec:theory_unitarity}. The DM {mass} then depends only on the dark gauge coupling $\alpha '$:
	\begin{eqnarray}
		m_{\rm dm}&\approx& 1.3\, \text{TeV}\,\left(\frac{0.1}{\alpha '}\right)\left(\frac{g'_{\ast s,\rm dec}}{\sqrt{g'_{\ast,\rm dec}}}\right)^{1/2}\left(\frac{30}{x'_{\rm dec}}\right)^{1/2}\label{eq:mass_t_channel}
	\end{eqnarray}
	{This gives the vertical dashed black line at $m_{\rm dm} \sim 2$ TeV in Fig. \ref{fig:polygon_t_ch_noDP}.}
	As explained in section \ref{sec:theory_unitarity}, the fate of the HS companions, here the $\gamma'$, may lead to a slight shift of position of the vertical line. Assuming that they decay into SM particles ({\em e.g.} through kinetic mixing) leads to the vertical line at $m_{\rm dm}\sim 2$~TeV obtained by imposing conservation of energy (instead of entropy, as in Eq.~(\ref{eq:mass_t_channel})), again see section \ref{sec:theory_unitarity}.

	{Note that }the dark photons can generically be made to decay before BBN, so that the domain {below the green shaded area but above} the green dashed line is potentially allowed. For the two sets of parameters considered here, where $m_{\gamma'}=10$~GeV and $m_{\gamma'}=30$~MeV, the constraint associated with the $\gamma'$, $\xi \lesssim m_{\gamma'}/T_{\rm BBN}$, is $\xi\lesssim 10^4$ and $\xi\lesssim 30$ respectively (or a more stringent constraint if the decay of the $\gamma'$ occurs after BBN has started, see e.g.~\cite{Berger:2016vxi,Hufnagel:2017dgo,Hufnagel:2018bjp}).
	The  gap in the mass range for candidates for which $\xi \gtrsim 1$ (see also Fig.~\ref{fig:polygon_t_ch}) could  be filled in several ways for this theory. Changing the coupling and the mass of the dark photon is one way. Another way, but a less effective one, is to play with the ratio of degrees of freedom between the VS and HS. We thus conclude that, with an appropriate choice of parameters (possibly all the way to the maximum cross sections allowed by unitarity), the DM candidates of this simple model can fill the whole domain of thermal DM candidates.

	\begin{figure}
		\centering
		\includegraphics[height=7.5cm]{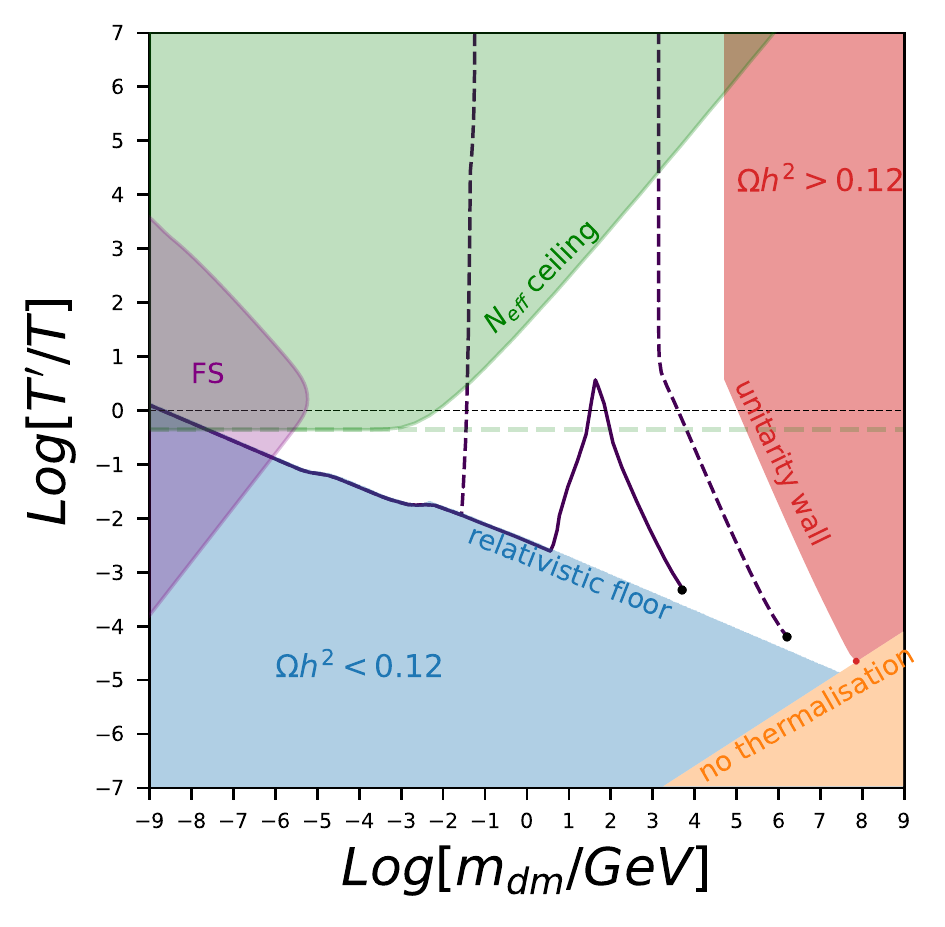}
		\caption{Parameter space for DM freeze-out driven by DM annihilating into two Dirac fermions in the s-channel mediated by a scalar. We used $\alpha_x=0.001$, $m_{\psi}=5$ GeV and $m_{\phi}=100$ GeV for the solid line and $\alpha_x=0.1$, $m_{\psi}=30$ MeV and $m_{\phi}=1$ GeV for the dashed line.}\label{fig:polygon_s_ch_noDP}
	\end{figure}
	
	\subsection{Scenario 2: s-channel}
	
	We next consider a model in which the DM can annihilate into a companion particle through a mediator in the s-channel. It is potentially richer than scenario 1 as it contains an extra coupling and more particles. Nevertheless, most of the features discussed in the case of scenario 1 are similar, so we will be brief. For definiteness, we consider the following Lagrangian,
	\begin{eqnarray}
		\mathcal{L}_{\rm HS} &=&  i \bar{\chi}\slashed{\partial}\chi - m_{\rm dm}\bar{\chi}\chi + i \bar{\psi}\slashed\partial\psi - m_{\psi}\bar{\psi}\psi\\
		&+& {\frac{1}{2}} \partial ^{\mu}\phi\partial _{\mu}\phi - {\frac{1}{2}} m_{\phi}^{2}\phi^{2} - y_{\chi}\phi\bar{\chi}\chi - y_{\psi}\phi\bar{\psi}\psi\nonumber
	\end{eqnarray}
	Here the DM is $\chi$, a Dirac fermion, and its companion particles are a scalar, $\phi$, and another Dirac fermion, $\psi$. If the $\psi$ is substantially heavier than the DM (and has a subleading contribution to the DM relic density or decays), we have essentially scenario 1, albeit with a spin zero particle instead of a dark photon. The new aspect is that the annihilation of $\chi$ into $\psi$, which is mediated by $\phi$, can be resonant and depends on $\alpha_x = y_\chi y_\psi/4\pi$. 
	
	Consider Fig.~\ref{fig:polygon_s_ch_noDP}, in particular, the solid curve, which mostly lies in the $\xi<1$ domain and for which $m_\phi = 100$ GeV, $m_\psi = 5$ GeV and $\alpha_x = 0.001$. The features of this curve are similar to those of the solid curve in Fig.~\ref{fig:polygon_t_ch_noDP} for the t-channel scenario. At low masses $m_{\rm dm} < m_\psi$, the DM candidates lie along the relativistic floor, then there is the threshold effect at $m_{\rm dm} \sim m_\psi$.  The most notable new feature  is due to resonant annihilation at $m_{\rm dm} \sim m_\phi$ which occurs here in the non-relativistic freeze-out regime. It peaks because the sharp drop of the DM abundance around the resonance (resulting from a sharp rise of the annihilation cross section) must be compensated by an increase of $\xi$. To the left of the resonance but after threshold, $\sigma v \sim \alpha_x^2 m_{\rm dm}^{2}/m_{\phi}^{4}$ so the curves grow as $\xi_{\rm dec} \propto \alpha_x^2 m_{\rm dm}^{2}/m_\phi^4$. To the right of the resonance, the mediator mass becomes less and less relevant and we recover the behaviour already observed in the t-channel scenario, $\xi_{\rm dec} \propto \alpha_x^2/m_{\rm dm}^{2}$. Again, the curve stops when the DM cannot thermalise.
	The features of the dashed line, obtained for a large coupling value, are similar to those of the dashed line in Fig.~\ref{fig:polygon_t_ch_noDP}. It shows a mass gap for the same reasons as discussed for that model. 
	
	\begin{figure}
		\centering
		\includegraphics[height=7.5cm]{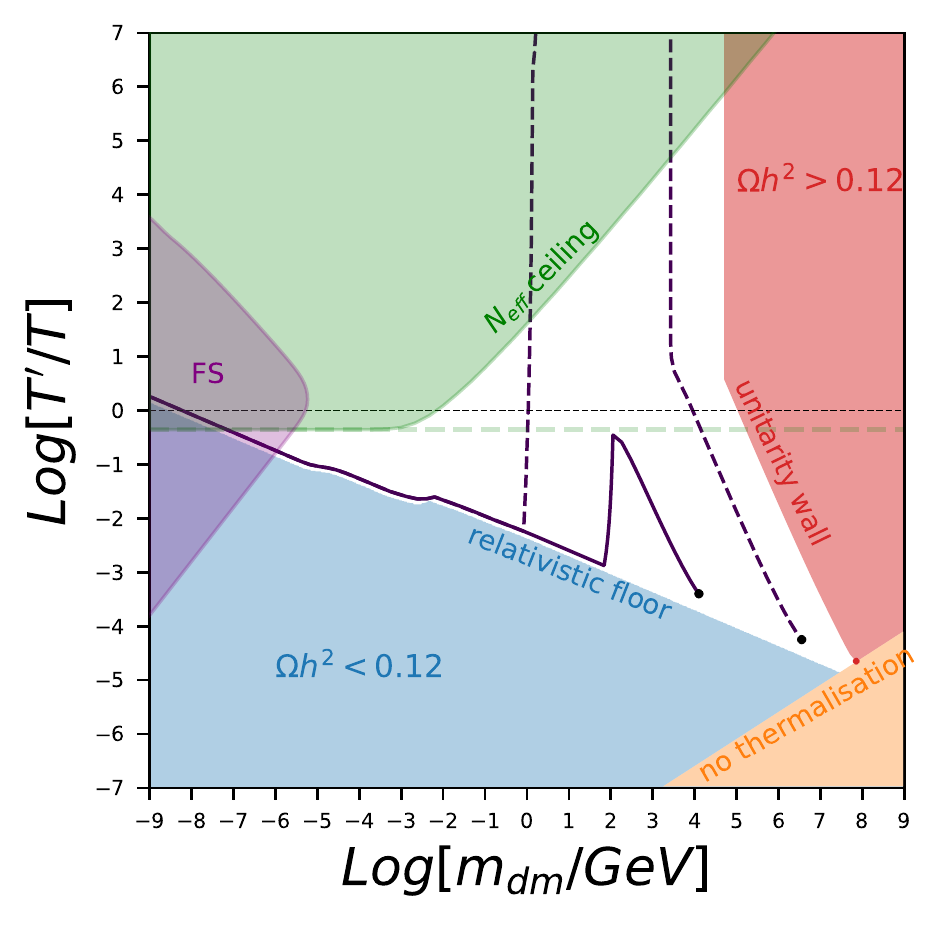}
		\caption{Parameter space for DM freeze-out driven by DM annihilation into two dark scalars via a contact interaction. {The} temperature ratio {is plotted} as a function of the DM candidate mass for $\lambda_{\Phi\eta}=0.01$ and $m_{\eta}=100$ GeV (solid line) and $\lambda_{\Phi\eta}=1$ and $m_{\eta}=1$ GeV (dashed line).}\label{fig:polygon_c_ch}
	\end{figure}
	
	\subsection{Scenario 3: contact interaction}
	Finally, we consider a model with no mediator, wherein the DM and companion particle interact via a contact interaction. 
	To that end, we introduce two real scalar fields, $\Phi$ and $\eta$, which are charged as $(+1,-1)$ and $(-1,+1)$ respectively under a $\mathbb{Z}_2 \times \mathbb{Z}_2$ symmetry. 
	The potential in the HS is then
	\begin{align}
		\mathcal{V}_{\rm HS} &= \frac{1}{2} m_{\rm dm}^2 \Phi^2 + {\lambda_\Phi\over 4!} \Phi^4  \notag \\
		&+ \frac{1}{2} m_\eta^2 \eta^2+ {\lambda_\eta\over  4!} \eta^4 + {\lambda_{\Phi\eta}\over 4} \Phi^2 \eta^2 \, .
	\end{align}
	For concreteness, we take the $\Phi$ as the DM candidate (thus the $\Phi$ field has no vev). Being scalars, they both could have portal couplings to the Higgs, $\vert H\vert^2 \Phi^2$ and $\vert H\vert^2\eta^2$. {As before, we assume that these are small enough that the HS does not thermalise with the VS}. The features of the $\xi(m_{\rm dm})$ contours are similar to the t-channel scenario, see Fig.~\ref{fig:polygon_c_ch}. First, there is a regime of relativistic freeze-out, then a sharp rise at the threshold {$m_{\rm dm} \approx m_\eta$} for $\Phi \Phi \rightarrow \eta \eta$ annihilation, and finally the $\xi_{\rm dec} \propto \lambda_{\Phi\eta}^2/m_{\rm dm}^{2}$ behaviour parallel to the unitarity wall. We depict these for the cases $m_\eta = 1$ GeV, $\lambda_{\Phi\eta} = 1$ {and $m_\eta = 100$ GeV, $\lambda_{\Phi\eta} = 0.01$}. 
	
	Despite these similarities, this simple model also 
	{allows us to consider the following question.
		So far we looked only at $2 \to 2$ annihilation processes. Could we go outside the domain by considering processes involving more particles? Although we will not look at this question in full generality, this model illustrates the fact that in general one can expect that the answer to this question is no. Besides the $2 \to 2$ processes, which can put
		$\eta$ and $\Phi$ particles in thermal equilibrium, 
		(with rate per unit volume $\Gamma_{2 \leftrightarrow 2}\propto \lambda_{i}^2 n_i^2/m_{\rm dm}^{2}$ with $i=\Phi,\eta$), the DM abundance can also}
	be changed by $4 \leftrightarrow 2$ processes, with $\Gamma_{4 \leftrightarrow 2}\propto \lambda_{i}^4 n_i^4{/m_{\rm dm}^{8}}$  (we follow the notation of \cite{Arcadi:2019oxh}, see also \cite{Bernal:2015xba}). Such processes are slower than the usual $2\leftrightarrow 2$ ones because of extra couplings, phase-space and, in the NR regime, Boltzmann factors. However, they {can nevertheless play 
		an important role.}

	First, we can imagine that the companion is simply absent (or heavier than the DM) in which case the HS consists only of the DM, $\Phi$. This is the scenario that was studied in detail in \cite{Arcadi:2019oxh}. 
	It is clear that FO cannot occur when the DM is relativistic, since $\Gamma_{4 \leftrightarrow 2} \propto T'$ for $T' \gg m_{\rm dm}$, rather FO occurs as the rate becomes Boltzmann suppressed for $m_{\rm dm} \lesssim T'$, see Fig.~\ref{fig:rates}. As there are no DM candidates along the relativistic floor, and no threshold from the companion, all DM candidates, for a given choice of self-coupling, lie (roughly) on a line that runs to the unitarity wall, but this refers to the unitarity of the $4\leftrightarrow 2$ process, not the DM freeze-out {from a companion}.
	The freeze-out of $4\leftrightarrow 2$ processes is distinct from that of the $2 \leftrightarrow 2$ case. According to the analysis of \cite{Arcadi:2019oxh}, the relic abundance is determined by  $\Gamma_{4 \rightarrow 2} \sim n_i H$. This condition leads to $\xi \propto (\lambda_{{\Phi}}/ m_{\rm dm})^{4/7}$, as opposed to $\xi \propto m_{\rm dm}^{-2}$ in the case of $2 \leftrightarrow 2$ processes, see Eq. (\ref{eq:UB_VS}). The slope is actually much closer to that of the relativistic floor, $\xi \propto m_{\rm dm}^{-1/3}$, see Eq. \eqref{eq:floor}. This stems from entropy conservation in the HS, which leads to reheating of the HS at the same time as the scalar particles deplete their number via self-interactions. The net effect is still a diminution of the particle abundance, but a much less drastic one than in the case of standard non-relativistic freeze-out. Still, the analysis of \cite{Arcadi:2019oxh} reveals that that cosmic DM abundance can be reached for a broad range of DM masses, with a maximum possible mass $m_{\rm dm} \sim 10^5$ GeV (see Fig.~8 in \cite{Arcadi:2019oxh}). Thus, DM candidates of this minimal scenario are well within the domain of thermal DM candidates.  They do not reach the unitarity wall (based on $2 \leftrightarrow 2$ processes), being qualitatively closer to the case of freeze-out in the relativistic regime.

	{A further interesting feature of this type of scenarios which is worth to point out is that} a similar mechanism could lead to the depletion of the companion particles themselves. 
	If $m_{\rm dm} \gg m_\eta$, such that the DM decouples before the $\eta$ self-interactions go out of equilibrium, then the situation is precisely the single scalar scenario studied by \cite{Arcadi:2019oxh}, except in this case the $\eta$ scalar is not DM and so its abundance should be subdominant. 
	Considering their Fig. 8 and fixing $m_\eta = 10$ MeV for concreteness, we can see that relatively small $\xi$ are required in order for the $\eta$ abundance not to be too large, and moreover there is a mild dependence on the quartic coupling. 
	Taking $\lambda_\eta = 10$ (about the largest allowed by unitarity), we find that $\Omega_\eta < \Omega_{\rm dm}$ for $\xi \lesssim 0.1$, while taking $\lambda_\eta = 10^{-3}$ (about the smallest that allows $\eta$ thermalisation), we have $\Omega_\eta < \Omega_{\rm dm}$ for $\xi \lesssim 0.03$. 
	Note that if $m_{\rm dm} \lesssim m_\eta$, the results of \cite{Arcadi:2019oxh} do not apply, indeed $\phi-\eta$ interactions could further suppress the $\eta$ abundance, thus in principle allowing larger $\xi$.
	These considerations only  apply to $\xi \lesssim 1$ since the entropy is conserved in the HS. If $\xi \gtrsim 1$, the DM must have companions {(as otherwise its abundance is too large)}, and they must decay back to the VS.
	
	\section{Conclusions}
	\label{sec:conclusions}
	Thermal relic dark matter candidates may come in many forms, and there is a vast literature concerning this class of models. 
	For scenarios in which the candidate was in equilibrium with the SM bath, the allowed range of dark matter masses is well known, as reviewed in \ref{sec:1D}. 
	It is, however, also plausible that dark matter thermalised within some hidden sector with a temperature, $T'$, different to the temperature of the SM bath, $T$. {This would happen for any hidden sector which involves relatively large interactions between the particles it contains but is connected to the SM thermal bath via significantly weaker interactions.}
	In this paper we studied this general scenario and identified the allowed domain of thermal dark matter candidates in terms of the DM mass and the ratio of temperatures, $\xi = T'/T$. 
	
	This domain is {given in  Fig.~\ref{fig:polygon}.
		While} {parts of} this result is implicit in many works, see \cite{Chu:2011be,Berlin:2016gtr,Arcadi:2019oxh,Hambye_2020,Hambye:2020lvy}, it provides a unifying, and to a large extent model-independent, picture. 
	In section \ref{sec:theory} we explored the theoretical bounds which lead to the exclusion of the blue (relativistic floor), red (unitarity wall) and orange (no thermalisation) regions depicted in the figure. 
	Moreover, trying to maintain the generality of our discussion, we  placed two observational bounds on the DM, as discussed in section \ref{sec:exp}. 
	This led to the green ($\Delta N_{\rm eff}$ ceiling) and purple (free streaming) exclusion regions in Fig.~\ref{fig:polygon}. 
	
	Putting everything together, we have identified the largest and smallest allowed mass and temperature of thermal DM candidates. 
	The DM mass range{, when it is a Dirac fermion,} is 
	\begin{equation}
		m_{\rm dm} \in [1.0 \text{ keV}, 52 \text{ PeV}] \, .
	\end{equation}
	The possible temperature {ratio} range is
	\begin{equation}
		\xi \in [1.4\cdot 10^{-5}, 6.9 \cdot 10^{5}]\,. 
	\end{equation}
	In particular, the lower right corner of the domain depicted in Fig. \ref{fig:polygon} corresponds to a candidate that decoupled while being mildly non-relativistic, $m_{\rm dm} \sim T'_{\rm fo}$. Hence, the corresponding temperature of the VS at the time of decoupling is bounded from above by $T_{\rm fo} \sim m_{\rm dm}/\xi \sim 10^{11}$ GeV. 
	
	Several other features, although rather obvious in retrospect, are made clear from our analysis. 
	Firstly, for $\xi < 1$, the permitted window of DM masses shrinks{, and shifts to larger values of $m_{\rm dm}$,} as $\xi$ decreases.
	Secondly, all other factors being kept constant, the function $\xi(m_{\rm dm})$ has in general an odd number of DM candidates, except for {very} fine-tuned instances.
	{These features are} illustrated by three simple models, discussed in section \ref{sec:models}. The results, plotted in Figs. \ref{fig:polygon_t_ch_noDP}, \ref{fig:polygon_t_ch}, \ref{fig:polygon_s_ch_noDP} and \ref{fig:polygon_c_ch}, give specific examples of the general findings of Fig.~\ref{fig:polygon}. 
	The different types of HS interactions considered\textemdash t-channel, s-channel and contact interaction\textemdash illustrate the applicability of our model-independent conclusions. The last model also includes scenarios in which the DM abundance is set by $4 \rightarrow 2$ processes.
	
	Several possible developments could be of interest.  First, we treated thermal decoupling with a broad brush. Although we do not expect our results to change significantly, it could be interesting to study more carefully and precisely how the DM abundance evolves if it is barely in thermal equilibrium, possibly in the vein of \cite{Bringmann:2020mgx}. Second, the observational constraint based on $N_{\rm eff}$ is very conservative and more stringent constraints, especially for candidates which decoupled (or have a companion that decayed back to the VS) around $T_{\rm BBN}$, could and should be derived. Another possibility is that some candidates have self-interactions that are constrained by {\em e.g.} the Bullet cluster \cite{Berlin:2016gtr}. All this is, however, model-dependent and beyond our scope. 
	
	Finally, we assumed that the portal interactions between the HS and the VS played little role in determining the relic abundance of the DM. Yet, they could be necessary to get rid of DM companions if their own abundance becomes a nuisance. This is particularly true for $\xi >1$ scenarios. We briefly mentioned the impact of such a connection if the mediator decays back to the SM, as for instance studied in \cite{Berlin:2016gtr}. One could also question how our picture changes if a portal interaction leads to a reannihilation regime \cite{Chu:2011be}. In this case, the HS thermalises while DM is being produced from the VS, and DM becomes non-relativistic when the production is still operative. However, one can check\footnote{Reannihilation regimes occur when interactions within the HS thermalise while the energy transfer from the VS to the HS is relevant but comparatively slow. Thus $\xi<1$ for reannihilation candidates. For unitarity limited interactions in the HS, from \cite{Cheung:2010gj,Chu:2011be} one can see that, for fixed $\xi <1$, the unitarity wall is  (slightly) shifted towards smaller $m_{\rm dm}$ and so the candidates lie within the thermal domain. Concretely, the relic density differs from the one obtained in the non-relativistic secluded freeze-out regime through a shift of $x'_f$ by $\ln ( [\langle \sigma v\rangle_{\rm portal}/\langle \sigma v\rangle_{\rm HS}]^{1/2}/\xi^2)$.} that DM candidates produced through reannihilation lie within the domain of thermal DM candidates.
	
	\section*{Acknowledgments}
	This work is supported by the F.R.S./FNRS under the Excellence of Science (EoS) project No. 30820817 - be.h ``The H boson gateway to physics beyond the Standard Model'', by the ``Probing dark matter with neutrinos" ULB-ARC convention, by the  FRIA, and by the IISN convention No. 4.4503.15. 
	R.C. thanks the UNSW School of Physics, where he is a Visiting Fellow, for their hospitality during part of this project.

	\bibliographystyle{apsrev}
	\bibliography{biblio}
	
\end{document}